\documentclass[journal=acscii,manuscript=article]{achemso}

\usepackage[version=3]{mhchem} % Formula subscripts using \ce{}
\usepackage[table]{xcolor}
\usepackage{caption}
\usepackage{subcaption}
\usepackage{multirow}
\usepackage{hyperref}

\newcommand{\UCB}{Department of Materials Science and Engineering, University of California, Berkeley, 210 Hearst Memorial Mining Building, Berkeley, CA, 94720, USA}
\newcommand{\LBL}{Materials Sciences Division, Lawrence Berkeley National Laboratory, 1 Cyclotron Road, Berkeley, CA, 94720, USA}
\newcommand{\ETA}{Energy Technologies Area, Lawrence Berkeley National Laboratory, 1 Cyclotron Road, Berkeley, CA, 94720, USA}
\newcommand{\TRI}{Present address: Toyota Research Institute, 4440 El Camino Real, Los Altos, CA, 94022, USA}

\author{Haoyan Huo}
\affiliation{\UCB}\alsoaffiliation{\LBL}

\author{Christopher J. Bartel}
\affiliation{\UCB}\alsoaffiliation{\LBL}

\author{Tanjin He}
\affiliation{\UCB}\alsoaffiliation{\LBL}

\author{Amalie Trewartha}
\affiliation{\LBL}\alsoaffiliation{\TRI}

\author{Alexander Dunn}
\affiliation{\UCB}\alsoaffiliation{\ETA}

\author{Bin Ouyang}
\affiliation{\UCB}\alsoaffiliation{\LBL}

\author{Anubhav Jain}
\affiliation{\ETA}

\author{Gerbrand Ceder}
\email{gceder@berkeley.edu}
\affiliation{\UCB}\alsoaffiliation{\LBL}

\title[ML Prediction of Solid-State Synthesis]
  {Machine-learning rationalization and prediction of solid-state synthesis conditions}

\abbreviations{ML,NLP,IR}

\begin{document}

%%%%%%%%%%%%%%%%%%%%%%%%%%%%%%%%%%%%%%%%%%%%%%%%%%%%%%%%%%%%%%%%%%%%%
%% The "tocentry" environment can be used to create an entry for the
%% graphical table of contents. It is given here as some journals
%% require that it is printed as part of the abstract page. It will
%% be automatically moved as appropriate.
%%%%%%%%%%%%%%%%%%%%%%%%%%%%%%%%%%%%%%%%%%%%%%%%%%%%%%%%%%%%%%%%%%%%%
\begin{tocentry}
\includegraphics[width=\textwidth]{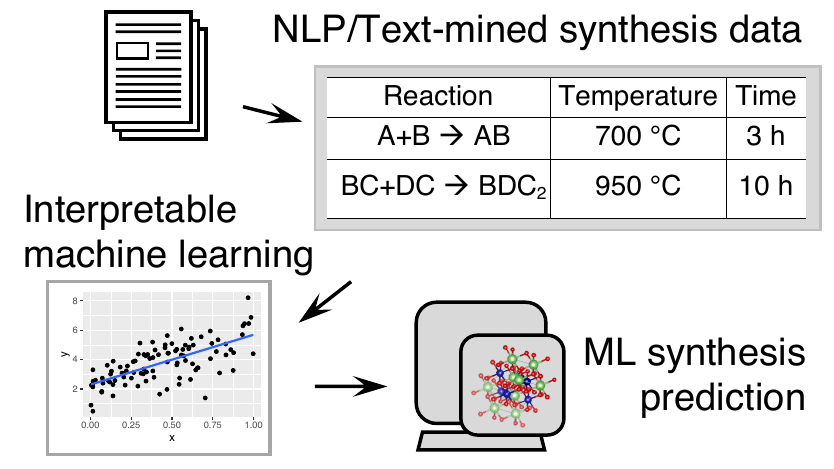}
\end{tocentry}

%%%%%%%%%%%%%%%%%%%%%%%%%%%%%%%%%%%%%%%%%%%%%%%%%%%%%%%%%%%%%%%%%%%%%
%% The abstract environment will automatically gobble the contents
%% if an abstract is not used by the target journal.
%%%%%%%%%%%%%%%%%%%%%%%%%%%%%%%%%%%%%%%%%%%%%%%%%%%%%%%%%%%%%%%%%%%%%
\begin{abstract}
% \textcolor{red}{Dated: Feb 15, 2022. Version 3, edit sequence: Tanjin He, Amalie Trewartha, Chris Bartel, Bin Ouyang, Chris Bartel, Gerd, Anubhav, Haoyan, Bin, Alex Dunn, Anubhav Jain, Tanjin He, Chris, Haoyan
% }

% Targeted journal: ACS Central Science: 14.553.
% Other journals to consider:
% Digital Discovery (RSC, new),
% Chemical Science 9.825,
% npj Computational Materials: 13.20,
% Chemistry of Materials: 9.811

% I also plan to submit the paper to arXiv after it's reasonably edited.

\vspace{5mm}

% \td{Please add your name. Thanks!}

There currently exist no quantitative methods to determine the appropriate conditions for solid-state synthesis. This not only hinders the experimental realization of novel materials but also complicates the interpretation and understanding of solid-state reaction mechanisms. Here, we demonstrate a machine-learning approach that predicts synthesis conditions using large solid-state synthesis datasets text-mined from scientific journal articles. Using feature importance ranking analysis, we discovered that optimal heating temperatures have strong correlations with the stability of precursor materials quantified using melting points and formation energies ($\Delta G_f$, $\Delta H_f$). In contrast, features derived from the thermodynamics of synthesis-related reactions did not directly correlate to the chosen heating temperatures. This correlation between optimal solid-state heating temperature and precursor stability extends Tamman's rule from intermetallics to oxide systems, suggesting the importance of reaction kinetics in determining synthesis conditions. Heating times are shown to be strongly correlated with the chosen experimental procedures and instrument setups, which may be indicative of human bias in the dataset. Using these predictive features, we constructed machine-learning models with good performance and general applicability to predict the conditions required to synthesize diverse chemical systems. 

\end{abstract}

%%%%%%%%%%%%%%%%%%%%%%%%%%%%%%%%%%%%%%%%%%%%%%%%%%%%%%%%%%%%%%%%%%%%%
%% Start the main part of the manuscript here.
%%%%%%%%%%%%%%%%%%%%%%%%%%%%%%%%%%%%%%%%%%%%%%%%%%%%%%%%%%%%%%%%%%%%%
\section{Introduction}

While solid-state synthesis is the prevailing approach for making inorganic solids, the determination of synthesis conditions for new solids is mostly based on heuristics and human-acquired experiences, with no analytical predictive approaches \cite{kohlmann2019looking,chamorro2018progress}. 
%  is a crucial step toward enabling the next generation of accelerated materials development \cite{szymanski2021toward,kimmig2021digital}. 
% Understanding why certain conditions are preferred for the synthesis of a given target material is a crucial step toward enabling the next generation of accelerated materials development \cite{szymanski2021toward,kimmig2021digital}. 
Recent work has focused on rationalizing solid-state reaction pathways observed in \textit{in-situ} experiments \cite{shoemaker2014situ,mcclain2021mechanistic,ito2021kinetically,paradis2020time,bianchini2020interplay}, by decomposing them into a sequence of phase evolution steps \cite{kohlmann2019looking} that can be modeled using thermodynamic calculations  \cite{miura2021observing,miura2020selective,mcdermott2021graph,aykol2021rational}. 
% With these approaches, solid-state synthesis has started to evolve from trial-and-error towards a synthesis-by-design paradigm  \cite{kovnir2021predictive,aykol2021rational,mcdermott2021graph}. 
To design synthesis routes for new materials, it is essential to understand why certain conditions are preferred and develop models for predicting these conditions for synthesis (e.g., temperature, time).
While thermodynamic calculations have been used to rationalize synthesis conditions in specific chemical systems \cite{miura2021observing,wustrow2021lowering}, a synthesis condition predictor with broad applicability for general inorganic compounds is still elusive.

% there are a few types of synthesis "models" (though model may be a bad word)

% 1) general models for helping design synthesis routes -- Aykol2021, McDermott2021
% 2) models for predicting whether something is synthesizable -- https://www.nature.com/articles/s43246-021-00219-x.pdf, https://pubs.acs.org/doi/abs/10.1021/jacs.0c07384
% 3) studies trying to understand how phases evolve during synthesis -- YBCO, NaMO2, etc

% this work stands out from (1) and (2) b/c it applies to synthesis conditions. this work stands out from (3) b/c it is broadly applicable (and applies to synthesis conditions)

% bartel.chrisj: not really "fitting"... i think there are two points that could be made to distinguish this work from these refs you cite: 1) most studies focus on a small number of targets (eg YBCO, MgCr2S4, NaMO2); 2) the "general" approaches focus on recommending precursors (Aykol 2021, Tanjin's paper?) or predicting the sequence of intermediates/reactions (McDermott 2021)

Here, we use statistical machine-learning (ML) methods to systematically learn and quantitatively evaluate synthesis condition predictors from a large set of experimental data. Such ML approaches require large, high-quality synthesis datasets covering many chemistries, which have only recently become available through the application of natural language processing (NLP) and information retrieval techniques on the large body of scientific literature \cite{kim2017machine,kim2017materials,kim2020inorganic,kononova2019text,vaucher2020automated,young2018data,karpovich2021inorganic}. In this work, using the dataset of over 30,000 text-mined solid-state synthesis reactions (denoted as the text-mined ``recipes" or the TMR dataset in this paper) \cite{kononova2019text}, we demonstrate an inductive ML approach that learns synthesis conditions from the knowledge parsed from the past literature.

\begin{figure*}
  \includegraphics[width=\textwidth]{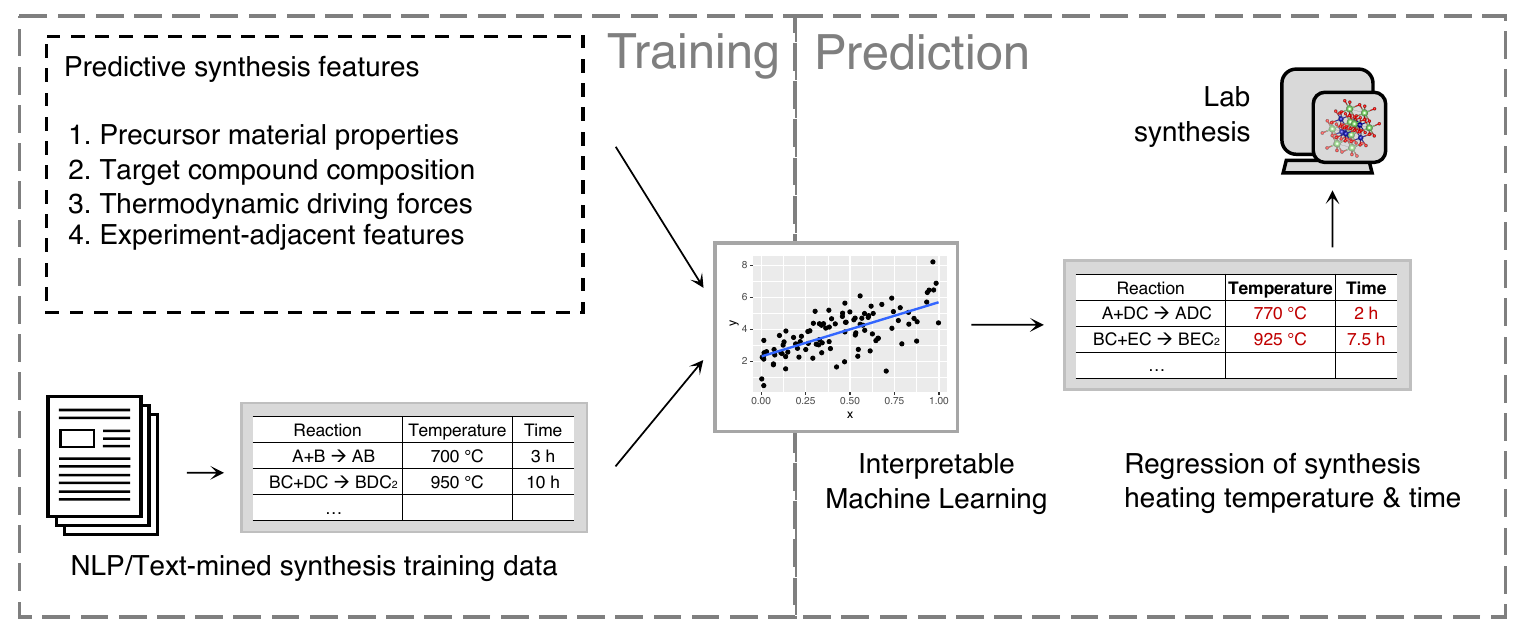}
  \caption{Schematic of the ML methods developed in this work for predicting solid-state synthesis conditions.}
  \label{fig:overview}
\end{figure*}

The overall pipeline of our ML approach is shown in Fig. \ref{fig:overview}. Datasets of synthesis conditions compiled from NLP/text-mined datasets are used to train ML models. Each synthesis reaction was represented using a set of human-designed features, which will be discussed in more detail in subsequent sections. Interpretable ML models were trained on this basis of features to predict two key solid-state synthesis conditions that must be specified for any reaction: heating temperature and heating time.

Throughout this paper, the prediction of solid-state synthesis conditions is defined as regression (point estimations) of the two experimental condition variables - temperature and time. Several important assumptions have been made: a) Good \textit{synthesizability} is assumed \cite{davariashtiyani2021predicting,sun2021generalized,sun2016thermodynamic,aykol2018thermodynamic}, i.e., when a publication reports the synthesis of some material at a specified set of conditions, we assume that this reaction was successful. b) Synthesis experiments are performed in a \textit{one-shot} fashion, i.e., reactants react and form the target compound in a single heating step, such that a simple synthesis route of ``mix and heat" would be sufficient. c) The ML models predict the ``optimal" synthesis conditions as implicitly defined by the consensus of training data. 

Note that the above assumptions oversimplify the synthesis condition prediction problem. These assumptions are often violated in many cases of practical solid-state syntheses. For example, a simple one-shot reaction route can thermodynamically favor an impurity phase which can only be avoided by using a multi-step synthesis with specific intermediate compounds \cite{dinia2004elaboration,aykol2021rational}; solid-state syntheses are often performed with many more degrees of freedom, such as special heating schedules \cite{dinia2004elaboration,miura2021observing}, special mixing devices \cite{shi2018shear}, different sintering aids \cite{rao2015essentials}, etc. Moreover, heating atmosphere strongly affects target material formation by changing the chemical potentials of gas species \cite{yuan2011different}. ML models require sufficient and consistent data to draw statistically significant conclusions \cite{montgomery2021introduction,friedman2017elements}, while the dataset used in this work has too imbalanced distributions for these additional labels. For example, only $<5\%$ of the reactions in the TMR dataset have non-air synthesis atmospheres. Therefore, the aforementioned conditions, although are present in the TMR dataset, are not predicted by the ML models in this work. Modeling of these factors may become possible as text-mined datasets become abundant in the future \cite{kononova2021opportunities}.

In this work, we considered 133 synthesis features describing four aspects of solid-state syntheses: 1) precursor properties, 2) composition of the target material, 3) reaction thermodynamics, and 4) experimental procedure setup. We ranked these features according to their predictive power using dominance importance (DI) analysis \cite{azen2003dominance}. The features were used to train linear and non-linear (tree-based) regressors for synthesis heating temperature and time. 
For all models, we split the dataset into reactions with carbonate precursors and reactions without carbonate reactions. This splitting is necessary because the release of CO2 gas in carbonate precursor materials systematically shifts the reaction driving forces for this subset and, consequently, the coefficients of the related features in linear models. Grouping the dataset into carbonate and non-carbonate reactions thus fits two sets of coefficients that accounts for this shift and improves the overall performance.
We performed leave-one-out cross-validation (LOOCV) to diagnose model performance. We also used out-of-sample (OOS) evaluation on Pearson's Crystal Data \cite{villars2021} (another synthesis dataset independently extracted from the literature, denoted as the PCD dataset in this paper) to test model generalizability on unseen datasets. The detailed data pre-processing and model construction can be found in the Methods section.

Our ML results achieve a goodness-of-fit measured by $R^2\sim 0.5-0.6$ and mean absolute error (MAE) $\sim 140 \mathrm{^\circ C}$ for heating temperature prediction. To compare with, typical heating temperatures used in solid-state synthesis range from $\sim 500 \mathrm{^\circ C}$ to $\sim 1500 \mathrm{^\circ}C$. For heating time prediction, the time variable is transformed into a new prediction variable representing reaction speed: $t \to \log_{10}(1/t)$. The goodness-of-fit for this new time variable is $R^2\sim 0.3$ and MAE is $\sim 0.3 \log_{10}(h^{-1})$ (e.g., if the predicted time is $t$, the MAE estimates a range of $[10^{-0.3}\cdot t, 10^{0.3}\cdot t]$, or $[0.5t, 2t]$). Analysis of the model predictive power reveals that heating temperature prediction is dominated by precursor properties, which we hypothesize to be linked to reaction kinetics. Heating time prediction is dominated by experimental operations, which may be indicative of human selection bias. The ML methods developed and applied in this work provide a statistically rigorous approach towards learning robust synthesis predictors from large datasets mined from the scientific literature.

\section{Results}

\subsection{Synthesis feature selection using dominance analysis}

In total, we created 133 features in four categories: 1) precursor properties - 12 features calculated from melting points, standard enthalpy of formation $\Delta H_f^{300K}$, and standard Gibbs free energy of formation $\Delta G_f^{300K}$ of precursors; 2) composition of the target material - 74 indicator variables representing the presence (1) or absence (0) of different chemical elements in the target compound; 3) reaction thermodynamics - 33 descriptive features of the driving forces for synthesis-relevant reactions constructed by decomposing synthesis into multi-step phase evolution paths using previously developed principles \cite{bianchini2020interplay,miura2021observing}; 4) experiment-adjacent features - 14 indicator variables representing whether certain devices, procedures, and/or additives were used in the synthesis procedure. See Methods for a more detailed description of how each of these classes of features were computed. 

\begin{figure*}[htbp]
  \includegraphics[width=\textwidth]{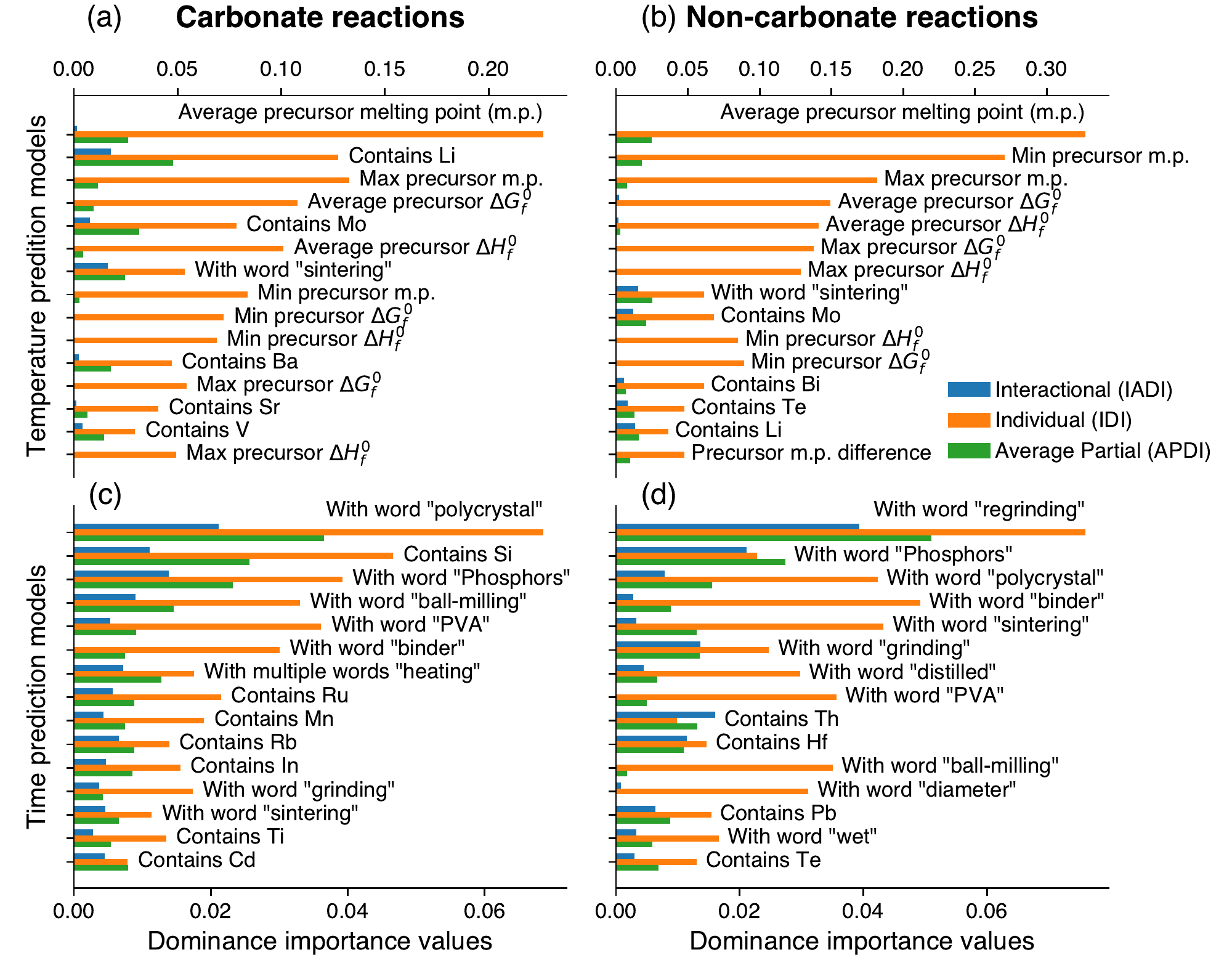}
  \caption{DI values and rankings of top 15 synthesis features for heating temperature models (a and b) and heating time models (c and d). The dataset is split into carbonate reactions (reactions with at least one carbonate precursor) (a and c) and non-carbonate reactions (reactions with no carbonate precursors) (b and d). Interactional dominance DI (IADI): decrease of model $R^2$ when a feature is removed from the whole model that uses all features. Individual dominance DI (IDI): $R^2$ of models trained using only one feature. Average partial dominance DI (APDI): average $R^2$ increase when a feature is added to a submodel. Features are ordered according to the sum of all three DI values.}
  \label{fig:di}
\end{figure*}

We first use DI analysis \cite{azen2003dominance} to rank the predictive power of these features. In DI analysis, one constructs many linear models that predict outcomes using subsets of features, called submodels. DI analysis then calculates the incremental effect of a feature $f_i$ on submodels that do not use $f_i$ in three different ways. 
% Specifically, we first identify those submodels not using $f_i$ as $\mathcal{F}(f_i) = \{F_j | F_j \cap \{f_i\} = \emptyset \}$ where $F_j$ denote subsets of features. 
The average partial dominance importance (APDI) value for $f_i$ is computed as the average increase of model performance, measured by $R^2$, when $f_i$ is added to any submodel that does not include $f_i$.
% : $\mathrm{E}_{\mathcal{F}}[ R^2(F_j \cup \{f_i\}) - R^2(F_j)]$, where $R^2(F)$ is the R-squared score of the linear model trained using features $F$. 
In other words, APDI measures the averaged gain of predictive power by including a feature. 
% Individual dominance importance (IDI) and interactional dominance importance (IADI) are two special cases of DI values. 
Individual dominance importance (IDI) values are the $R^2$ of models trained using only one feature and quantify the predictive power of the features by themselves. Interactional dominance importance (IADI) values are the decrease of model $R^2$ when a feature is removed from the whole model that uses all features, therefore measuring the gain of predictive power by a feature over all other features. All three DI values are computed for both heating temperature and time prediction models and are shown in Fig. \ref{fig:di}.
We split the dataset into carbonate reactions (reactions with at least one carbonate precursor) and non-carbonate reactions (reactions with no carbonate precursors). This is necessary because these two subsets have dissimilar distributions of reaction thermodynamic driving forces, which must be separated to be modeled in linear regression \cite{faria2010fitting,li2018learning}. 

We first evaluate the predictive powers of the features by themselves, as demonstrated by the IDI values in Fig. \ref{fig:di}. For heating temperature prediction, Fig. \ref{fig:di} (a) and (b) show that the IDI values of the average precursor melting points are significantly higher than those of other features. Average precursor melting points alone achieve $R^2\sim 0.2 - 0.3$ for heating temperature prediction. Other features, such as experimental Gibbs free energy of formation at standard conditions $\Delta G_f^{300K}$ and experimental enthalpy of formation at standard conditions $\Delta H_f^{300K}$ of precursors, are also highly predictive features as measured by IDI. Note that precursor melting points, $\Delta G_f^{300K}$, and $\Delta H_f^{300K}$ are likely to be good proxy variables for precursor reactivity. 
% Chris:
%  i think dHf/dGf_precursors implying kinetics is only valid taken together with the observation that dG_reaction is not very predictive. 
% If both dGf_precursor and dG_reaction were important features, i would be inclined to argue that this is telling us thermodynamics is important and dGf_precursor is just predicting dGf_reaction
% We therefore hypothesize that the optimal heating temperature is highly dependent on reaction kinetics, for which precursor reactivity plays an important role. 
The next set of predictive features as ranked by IDI are compositional indicator variables (e.g., indicating the presence/absence of \ce{Li}, \ce{Mo}, \ce{Bi}, etc.). These features can be understood as chemistry-specific corrections to heating temperatures. Note that ML models aim to reduce prediction errors for the whole training dataset, which is dominated by the elements that are characteristic of large application fields, such as \ce{Li} (Li-ion batteries) and \ce{Ba} (perovskite oxides). It is thus not surprising that these most frequently synthesized chemical systems appear at the top of the list in Fig. \ref{fig:di} (a) and (b).

For heating time prediction, Fig. \ref{fig:di} (c) and (d) show that the IDI of experiment-adjacent features (e.g., indicators of polycrystal synthesis, phosphors, and usage of ball-milling devices) completely outweigh precursor property features. This suggests that heating time is largely controlled by the desired applications (e.g., the need for dense pellets, small particles, single crystals, etc.) and experimental setups rather than reaction mechanisms. 
% Intuitively, we may also think that heating time is often chosen based on convenience of preparing and extracting samples during working hours and therefore not fully optimized. 
Meanwhile, compositional indicator variables still rank second after the experiment-adjacent features, again acting as chemistry-specific corrections.

The blue bars in Fig. \ref{fig:di} are IADI values. IADI values measure the gain of predictive power by a feature over all other features. For heating temperature prediction, Fig. \ref{fig:di} (a) and (b) show that IADI values are very small for most features. A low IADI value is usually due to high correlation among features, e.g., average precursor melting points and maximal precursor melting points. These high correlations suggest it is necessary to use feature selection to choose the strongest feature among highly correlated features, as will be discussed in the next section. Nevertheless, a few features have relatively higher IADI values, a sign that they bring unique extra information over all other features. For example, describing syntheses using the word ``sintering" may suggest the experimenters actively chose higher heating temperatures. As a consequence, the experiment-adjacent feature of ``sintering" has the highest IADI value for temperature prediction models.

The green bars in Fig. \ref{fig:di} are APDI values. APDI values are the average $R^2$ increase of a feature to all submodels. Thus, APDI estimates the general usefulness of a feature. APDI and IDI values are therefore two important factors in ranking feature importance. For example, in Fig. \ref{fig:di} (a), even though average precursor melting point and $\Delta G_f^{300K}$ both have high IDI values, $\Delta G_f^{300K}$ has smaller APDI values and is less important due to correlation with alternative features. By ranking all features according to the summation of DI values, we are able to consistently select the most uniquely predictive features. 

While in general, synthesis temperature and time together determine the overall reaction kinetics, they are not ranked as top predictive features in Fig. \ref{fig:di} when included as features to predict each other (also see Table S1). This seems contrary to the expectation that they would be strongly correlated because elevated temperatures can lead to faster reactions by promoting atomic diffusion. We hypothesize that the low correlation between time and temperature may be due to a variety of reasons:
1) As opposed to sampling many synthesis conditions for a specific chemical system, the TMR dataset spans diverse chemistries. There are usually less than 5 reported syntheses for a majority ($>60\%$) of the chemical systems which is not enough to reveal a stronger correlation, and 2) The TMR dataset is text-mined from journal articles in which synthesis conditions, especially synthesis time, are generally not optimized but are determined by other external factors, such as the desired applications or the researcher's convenience. These external factors make the time variable more noisy and less correlated to temperature than it might be in a variationally constrained set of data (e.g., the collection of shortest times for each temperature)

To summarize, the overall rankings in Fig. \ref{fig:di} suggest each prediction variable is dominated by two types of features. For heating temperature prediction, precursor material properties have the most feature importance, while compositional features act as secondary corrections. For heating time prediction, experiment-adjacent features dominate the prediction, while compositional features also provide secondary corrections. Contrary to the common application of decomposing synthesis reactions into multi-step phase evolution paths using thermodynamic principles \cite{mcdermott2021graph,aykol2021rational,wustrow2021lowering,miura2021observing}, Fig. \ref{fig:di} shows the phase evolution thermodynamic driving force features, developed using similar principles in this work, provide little predictive power for heating temperature and time. 
% This surprising result suggests thermodynamics alone is insufficient to explain the synthesis conditions reported in the literature. 
We hypothesize that this is due to the fact that the TMR dataset contains only positive experimental results for which researchers actively optimize for reasonable reaction kinetics. Therefore, reaction driving forces are less useful as these features are more likely to indicate whether something is synthesizable (e.g., if reactions to form a target are thermodynamically spontaneous) rather than indicate at what conditions reactions may occur quickly. We will revisit this finding in more detail in the Discussion section. 

\subsection{Building and interpreting linear regression models}

\begin{figure*}[htbp]
  \includegraphics[width=\textwidth]{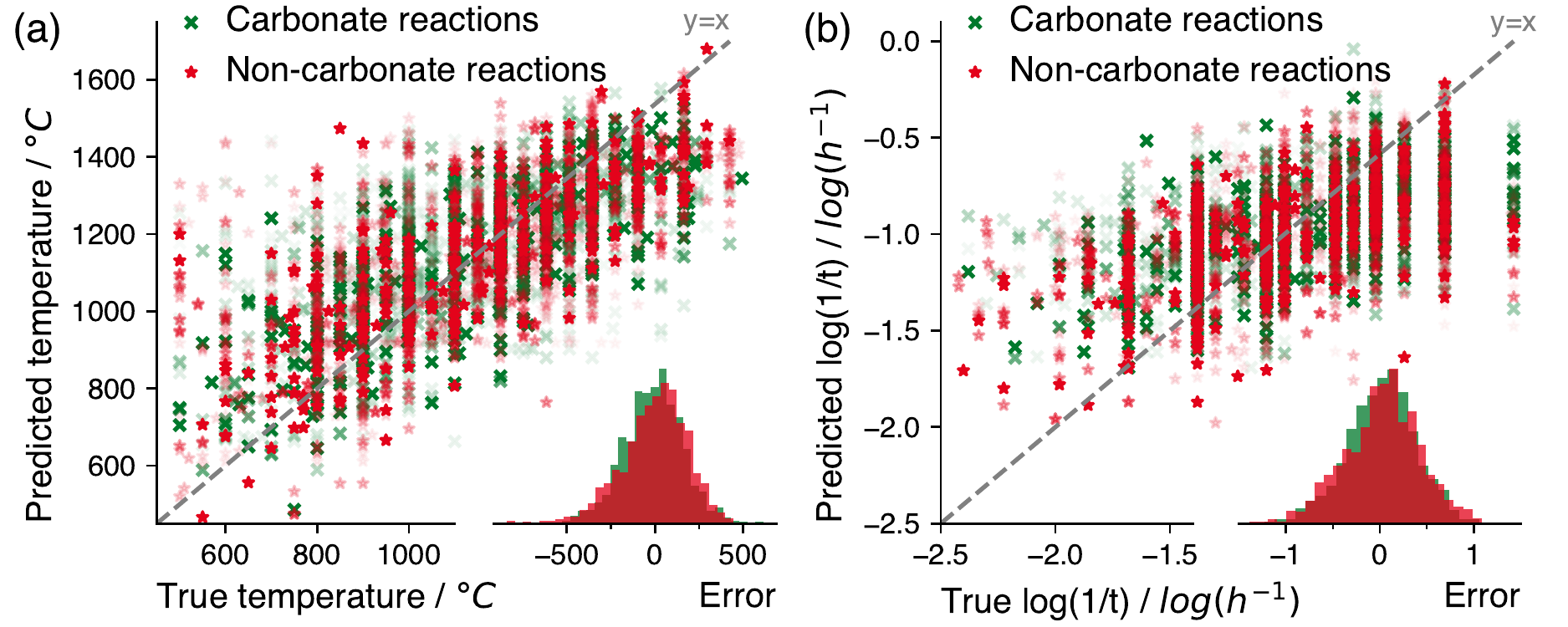}
  \caption{Regression result of linear models. The scatter plots show reported conditions v.s. predicted conditions for temperature prediction (a) and time prediction (b). Opacity of the markers indicates the weights of data points. Histograms of prediction errors are also shown.}
  \label{fig:full-model}
\end{figure*}

To build regression models, we start with linear regressors as baseline models since their good interpretability allows one to focus on feature engineering and decipher the relations between features and synthesis conditions. To balance between high predictive power and possible overfitting, we add features in the order of DI rankings and drop any feature that increases model Bayesian information criterion (BIC) values \cite{friedman2017elements}. In total, four linear models (heating temperature and time prediction models for carbonate and non-carbonate reactions) were trained using weighted least squares (WLS) \cite{friedman2017elements}. The scatter plots of the predicted synthesis conditions versus the reported conditions are shown in Fig. \ref{fig:full-model} (a) and (b). For heating temperature prediction, the $R^2$ values of the models are 0.55 on carbonate reactions and 0.56 on non-carbonate reactions, while the MAE are $134 \: \mathrm{^\circ C}$ and $147 \: \mathrm{^\circ C}$, respectively. For heating time prediction, the $R^2$ values of the models are 0.31 on carbonate reactions and 0.33 on non-carbonate reactions, while the MAE are $0.30 \log_{10}(h^{-1})$ and $0.32 \log_{10}(h^{-1})$, respectively. Since we predict the transformed time variable $\log_{10}(1/t)$, such MAE estimates the time prediction is within range $[10^
{-0.3} \cdot t, 10^{0.3}\cdot t]$, or $[0.5t, 2t]$ (e.g., for a 2-hour experiment, the expected prediction range is $0.5-4$ hours). Note that these metrics are evaluated on training data. Thus, they may not reflect the model performance when applied on unseen data. We will perform cross validation and discuss the results in later sections.

\begin{figure*}[htbp]
  \includegraphics[width=\textwidth]{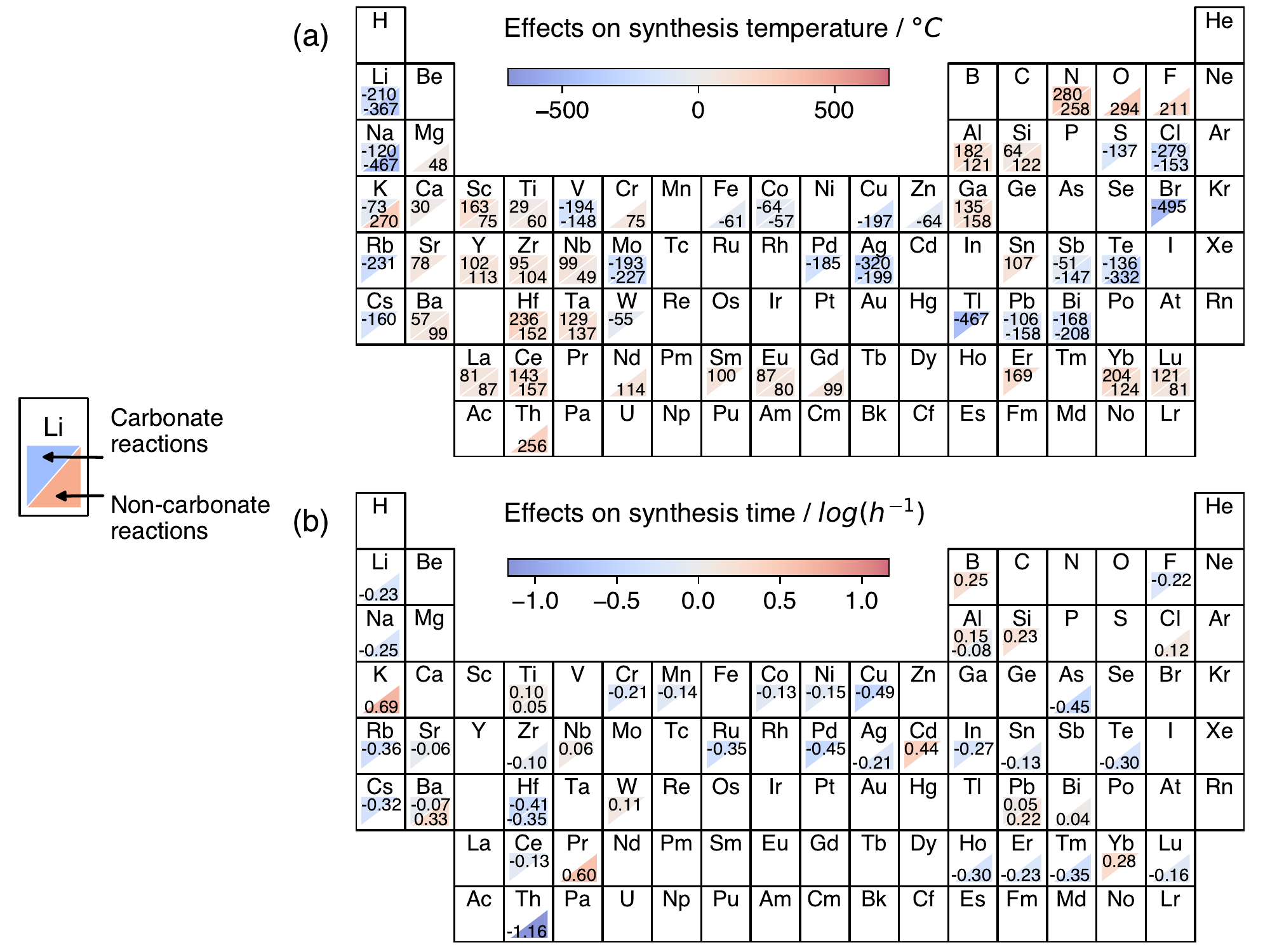}
  \caption{The average effect of each chemical element to predicted heating temperatures (a) and times (b) in trained linear models. The values are coefficients of the corresponding features in the linear models, quantifying how much the predicted value changes relatively if a new chemical element is added to (or removed from) the synthesis.}
  \label{fig:elemental-effects}
\end{figure*}

In a linear regressor $\hat{y} = \sum_i \beta_i x_i$, the feature coefficients $\beta_i$ quantify how the regression target variable responds to unit changes of $x_i$. As a special case, when $x_i \in \{0, 1\}$ are indicator variables (e.g., compositional and experimental-adjacent features), $\beta_i$ can be interpreted as additive effects on the prediction target variable when features $x_i = 1$. For all compositional features, the effects are shown in Fig. \ref{fig:elemental-effects} (a) and (b). Note that these values are relative to the ``average" according to the training dataset and must be interpreted in relative values. For example, if \ce{Li} is present in the target compound, Fig. \ref{fig:elemental-effects} (a) suggests the heating temperature will decrease by $360\: \mathrm{^\circ C}$ on average for non-carbonate reactions. On the other hand, the presence of \ce{N} will increase the heating temperature by $260\:\mathrm{^\circ C}$ on average. Therefore, Fig. \ref{fig:elemental-effects} (a) and (b) are maps that associate different chemistries with their effect on optimal synthesis conditions. Such maps can be used as empirical ``synthesis rules” that are helpful for designing synthesis routes to new materials. % For example, Fig. \ref{fig:elemental-effects} (a) suggests adopting a synthesis for \ce{LiFePO4} for synthesizing \ce{NaFePO4} would require a higher temperature of $\approx 100 ^\circ C$ on average.

The learned coefficients in Fig. \ref{fig:elemental-effects} (a) and (b) are sparse because some elements appear only a few times or are even missing in the training dataset, precluding a confident estimate of their effect (assessed by the p-values of the coefficients with a $5\%$ significance level \cite{seabold2010statsmodels}). In Fig. \ref{fig:elemental-effects}, we observe more consistent compositional effects across similar element periods and groups for temperature predictions than for heating time predictions. The lack of correlation with  compositional effects for time prediction matches the DI analysis result in Fig. \ref{fig:di} (c) and (d), which suggests compositional features are less helpful for predicting heating time. 
% Therefore, compositional features are more likely to capture bias towards particular data points for heating time prediction. 
Moreover, the compositional effects are less consistent between carbonate reactions and non-carbonate reactions for heating time prediction. These observations suggests the compositional effects are generally less reliable for heating time prediction and must be used with more caution.

\subsection{Training and cross-validating non-linear models}

\begin{figure*}[ht]
  \includegraphics[width=\textwidth]{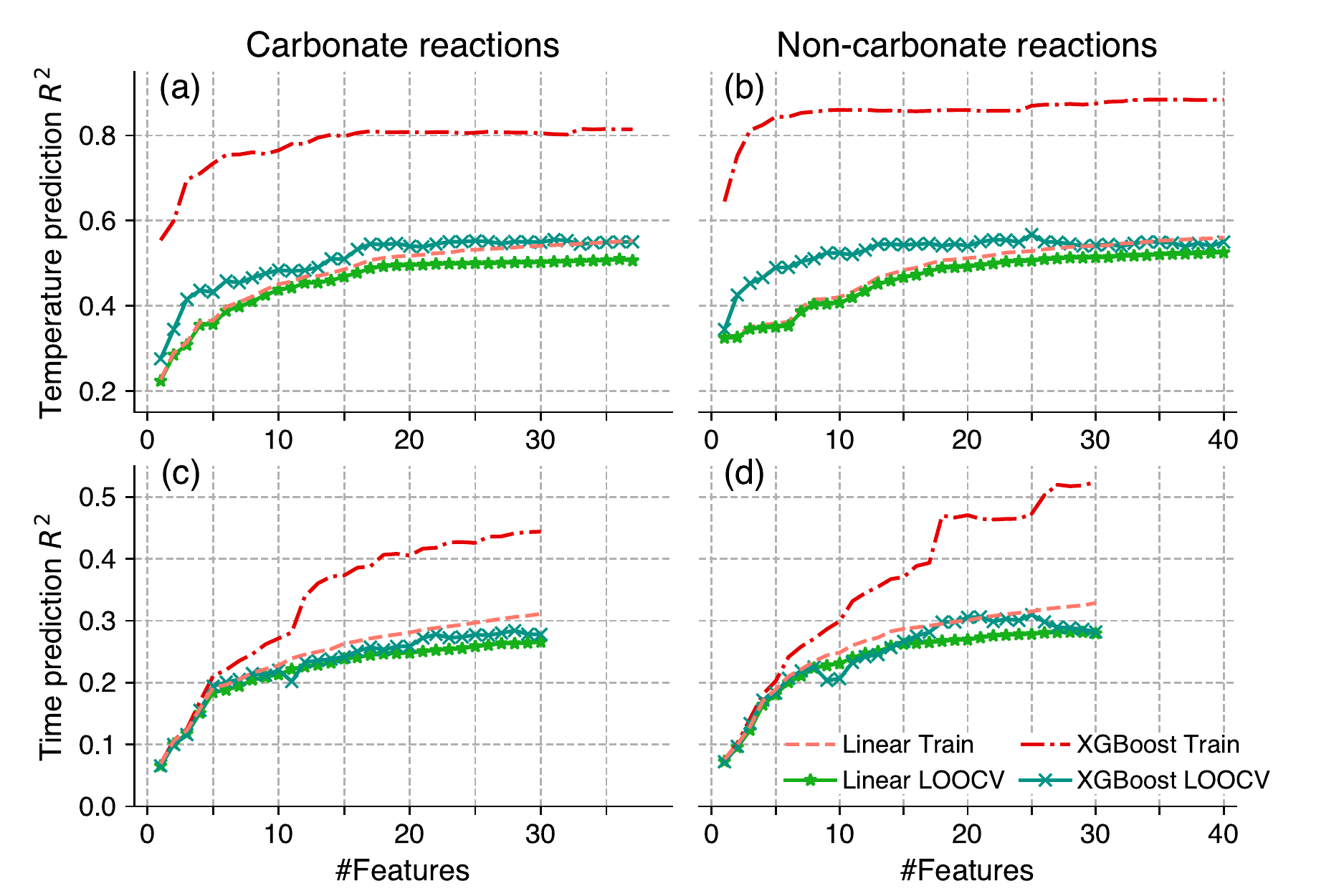}
  \caption{Model performance versus number of training features for both linear and non-linear (gradient boosting tree regressor) models. The x-axis shows the number of features used. The features are added in the order of DI value rankings. The first row shows performances of temperature prediction models trained on carbonate reactions (a) and non-carbonate reactions (b). The second row shows performances of time prediction models trained on reactions with (c) and without (d) carbonate precursors.}
  \label{fig:non-linear-models}
\end{figure*}

Having used DI analysis and linear models to probe the synthesis prediction features, we next aim to systematically cross-validate ML models to understand their generalizability or propensity for overfitting. Fig. \ref{fig:non-linear-models} shows the model performances versus the number of features, which characterize training $R^2$ and the LOOCV Pseudo-$R^2$ (a metric comparable to $R^2$, see Methods ) scores of the linear models as more features are included in training. In Fig. \ref{fig:non-linear-models}, features are added into the models in the order of DI value rankings. Fig. \ref{fig:non-linear-models} shows that both training and LOOCV scores increase quickly when the number of features is less than 10. This result is consistent with the DI values in Fig. \ref{fig:di} as the first few features have the highest feature importance. The model performance continues to improve as we include all other features, although the marginal improvement decreases rapidly. The training and LOOCV curves for linear models exhibit very similar performance, suggesting that these linear models have little risk of overfitting.

\begin{figure*}[ht]
  \includegraphics[width=\textwidth]{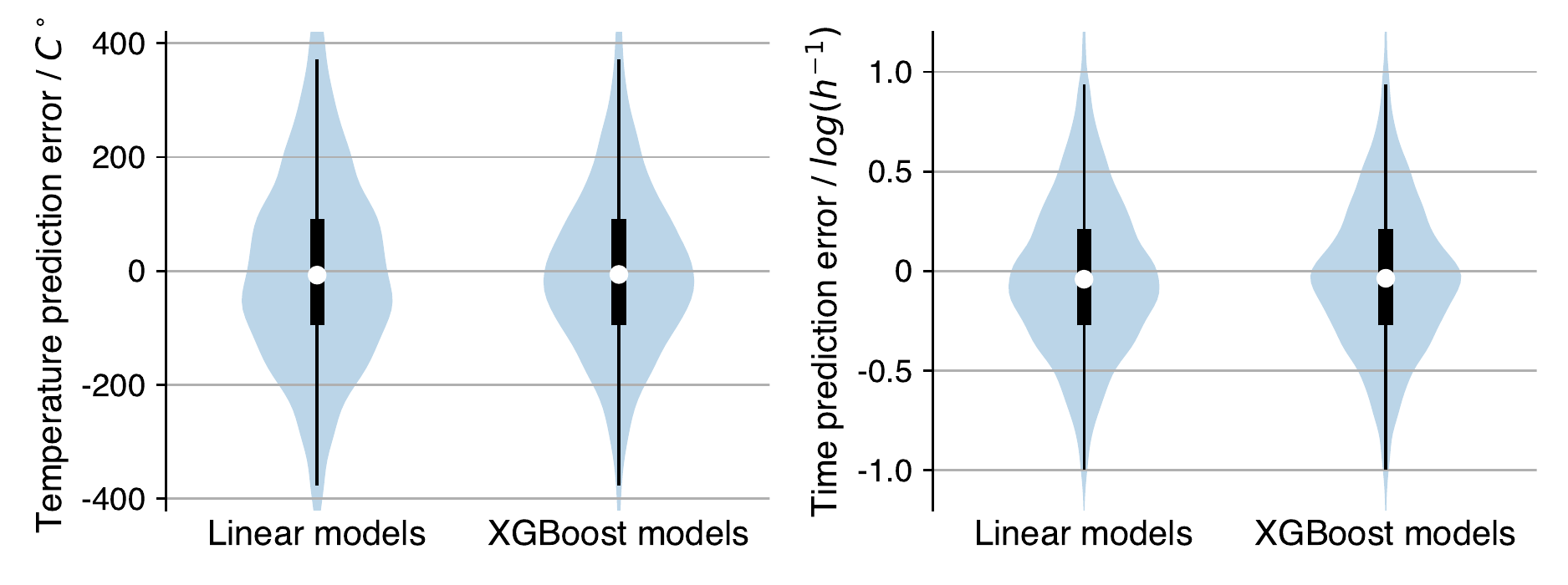}
  \caption{LOOCV prediction error distributions of synthesis temperature and time. Plotted are prediction median values (shown with white dots), interquartile ranges (IQR, or the spread of errors between 25\% and 75\% percentiles, shown with thick lines), and 1.5x IQR (shown with thin lines). Shaded areas are probabilistic density estimations of the errors. Our models are expected to make prediction errors within the IQR half of the times and within the 1.5x IQR most of the times.}
  \label{fig:violin-error}
\end{figure*}

The linear model may be incapable of capturing non-linear correlations among features and synthesis conditions. We next use advanced ML models that are capable of modeling non-linear relations on the same set of features as for the linear models. Among many ML models we attempted during preliminary experiments, gradient boosted regression trees (GBRT), implemented in the XGBoost package \cite{chen2016xgboost}, demonstrated the best LOOCV scores after proper hyperparameter tuning. XGBoost models use a large number of weak tree learners to build a strong ensemble regressor and are able to learn non-linear effects. Indeed, we observe in Fig. \ref{fig:non-linear-models} that XGBoost training Pseudo-$R^2$ (red dashed curves) are significantly higher than linear models. However, as shown by the teal crosses in Fig. \ref{fig:non-linear-models}, compared to the LOOCV scores of linear models (green stars), the LOOCV Pseudo-$R^2$ scores of XGBoost models do not improve as much when compared to the LOOCV performance of the linear models, suggesting an increased level of overfitting by XGBoost models. One advantage of XGBoost over linear models is improved utilization of a small number of features, as shown by the steeper curves when the number of features is less than 10 in Fig. \ref{fig:non-linear-models} (a) and (b), although the advantage diminishes once sufficiently many features are used. Finally, to help better understand the uncertainties of the models, we visualize the error distributions of synthesis conditions in Fig. \ref{fig:violin-error} using violin plots, where we mark the interquartile range (IQR) representing 50\% of the errors, and 1.5x IQR, representing the range of prediction errors beyond which are considered outliers.

% In a real-world application, the error rate of a ML model applied on unseen data can usually be decomposed into three parts: $\mathrm{model\: bias} + \mathrm{model\: variance} + \mathrm{dataset\: shift}$. As shown by the curves in Fig. \ref{fig:non-linear-models}, linear models have higher training error than XGBoost models (larger model bias), but XGBoost models have higher gap between training and LOOCV errors (larger model variance). The combined model bias and variance 

\subsection{Testing model generalizability using the PCD dataset}

\begin{figure*}[ht]
  \includegraphics[width=\textwidth]{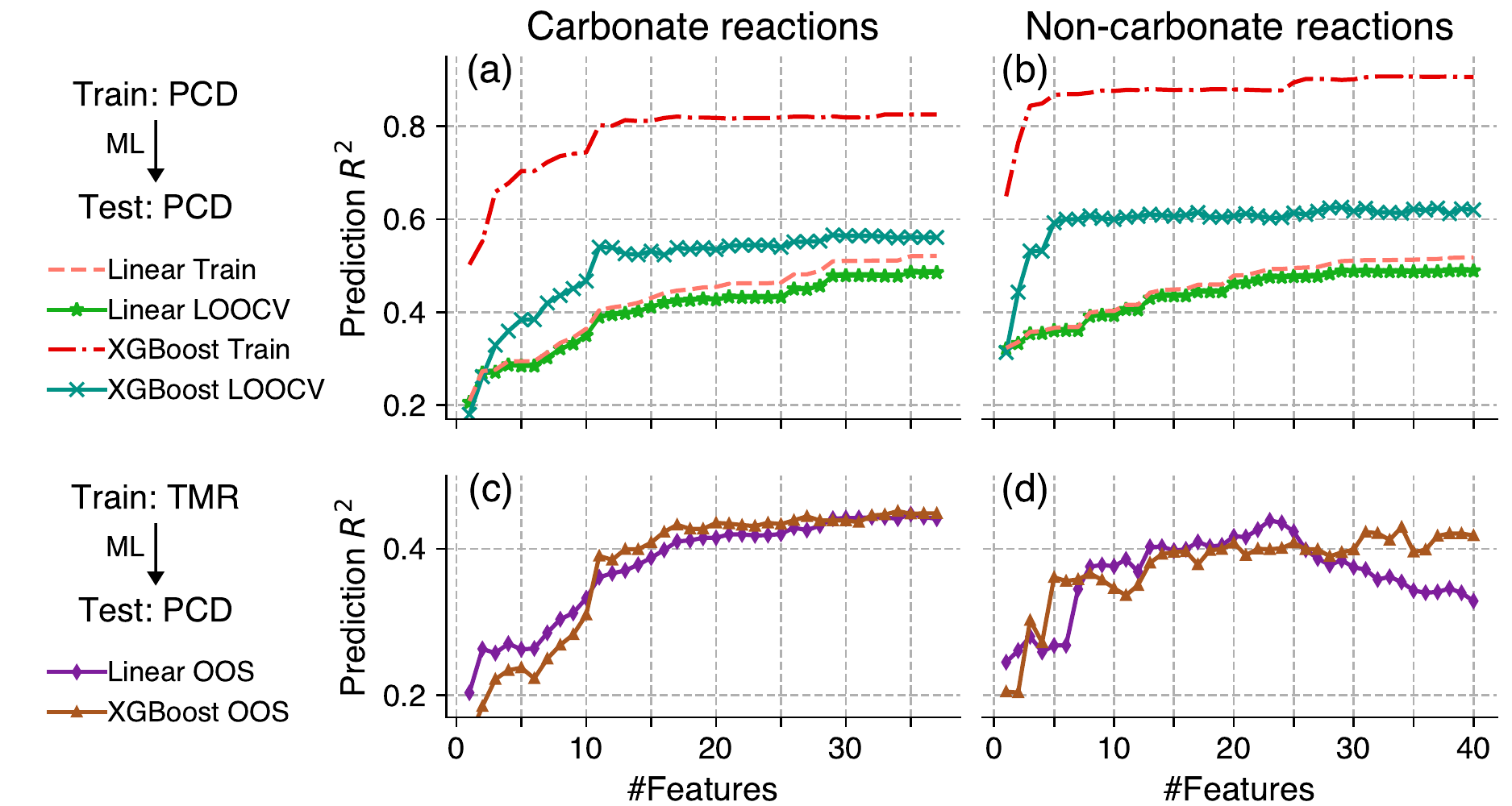}
  \caption{Performance of the models versus the number of features evaluated on the PCD dataset. X-axes show the number of features used in each model. Features are added in the order of DI value rankings as in Fig. \ref{fig:di}. The left panels (a) and (c) show models trained on carbonate reactions and the right panels (b) and (d) show models trained on non-carbonate reactions. Top panels (a) and (b) show performance of models trained and evaluated on the PCD dataset, which represent the upper bounds of OOS scores (c) and (d), which show performance of the models trained on the TMR dataset. A higher OOS score indicate better model generalizability. }
  \label{fig:oos}
\end{figure*}

When applied to unseen datasets, ML model predictions tend to have larger errors due to dataset shift, i.e., unseen datasets have a different distribution than the training datasets \cite{quinonero2009dataset}. 
%The TMR dataset is text-mined from a general collection of inorganic chemistry \cite{kononova2019text}, but the ML models may be applied to, for example, electrochemical materials, leading to \textit{covariate shift} in the distribution of target compound chemistries \cite{quinonero2009dataset}. 
In particular, the relations between features and outcomes may change for unseen data, leading to \textit{concept drift}, degrading model generalizability and limiting model applicability.

The TMR dataset mostly contains syntheses for inorganic oxide materials and is dominated by target materials containing Ti, Sr, Li, Ba, La, Nb, Fe, etc., reflecting popular materials in the inorganic materials research community such as perovskite oxides and battery materials. The TMR dataset also contains a large fraction of solid solutions or doped materials. To estimate and understand how the ML model trained on the TMR dataset generalizes to unseen datasets, we utilized the PCD dataset as an additional test. The original PCD collection contains inorganic materials syntheses that were manually extracted from the literature in a semi-structured natural language form \cite{villars2021}. We processed the PCD (Pearson's Crystal Data) collection using the same text-mining pipeline and only kept oxide syntheses such that the final PCD dataset has a similar chemistry distribution as the TMR dataset. To ensure there are no duplicate syntheses, we removed any entry in the PCD dataset whose digital object identifier (DOI) is present in the TMR dataset (i.e., syntheses in same papers are not allowed, but the same compositions from different papers are allowed). Compared to the TMR dataset, the PCD dataset shares a similar distribution of chemical systems and synthesis conditions, as indicated by similar sets of popular chemical elements (i.e. Ti, Fe, Sr, Ba, Si, etc.) and average synthesis temperatures around $1200^\circ C$, see Fig. S3. The PCD dataset thus represents a reasonable benchmark dataset for our ML models. However, since many reactions in the PCD dataset do not have heating times extracted, we only predicted heating temperatures for the PCD dataset. 

To establish an upper bound of the model performance, we performed the same training/validation procedure using the PCD dataset as was used on the TMR dataset. Fig. \ref{fig:oos} shows the performance of the ML models versus the number of features. The green stars and teal crosses in Fig. \ref{fig:oos} are the LOOCV scores of linear and XGBoost models, respectively. XGBoost models achieve $0.5 \sim 0.6$ LOOCV Pseudo-$R^2$ which is considerably better than linear models ($0.4\sim 0.5$). Moreover, XGBoost shows steeper performance increase when few synthesis features are used. Compared to Fig. \ref{fig:non-linear-models}, the advantage of the non-linear models are much more substantial for the PCD dataset than for the TMR dataset. This clear advantage of XGBoost models indicates they are more robust than linear models against possible dataset shift effects.

Next, we performed tests to understand how well ML models trained on the TMR dataset are generalizable to the PCD dataset. The purple diamonds and yellow-brown triangles in Fig. \ref{fig:oos} show the OOS performances of the linear and XGBoost models trained using the TMR dataset but evaluated on the PCD dataset. It is interesting to note that XGBoost and linear models have very similar OOS scores for carbonate reactions, but XGBoost clearly outperforms linear models for non-carbonate reactions when more ($>30$) features are used. Upon further investigation, the features \#30 to \#40 used on non-carbonate reactions are mostly related to thermodynamic properties of the reactions. The performance drop after features \#30 suggests that relations between thermodynamic features and heating temperatures learned on the TMR dataset by linear models do not transfer well to the PCD dataset. On the other hand, XGBoost models seem to be able to consistently maintain good performance regardless of the number of features used.

% Some key points to consider here:

% Why would one want to test using a external dataset? 
% A: to address dataset shift as it is very common in applied ML in physical sciences.

% What have we found?
% A: First, there is indeed performance degradation caused by dataset shift. Second, XGBoost is more robust against dataset shift.

% What are the implications?
% A: retrain model periodically. Understanding the underlying dataset distribution is important. XGBoost has a better genralizability.

In Fig. \ref{fig:oos}, the difference between LOOCV scores and OOS scores confirms the ML models have degraded prediction performance ($R^2$ drops by 0.1) when applied to a different dataset. The performance degradation caused by dataset shift is often inevitable and requires regularly retraining the ML models in order to adapt to the new datasets. However, Fig. \ref{fig:oos} suggests XGBoost models are more robust against dataset shift and have a better generalizability. We hypothesize this is due to the strong regularization and therefore recommend ML synthesis condition predictors to be built with XGBoost or similarly regularized models.

\section{Discussion}

ML predictions must be statistically evaluated using large datasets, so this work has focused heavily on reducing the expected prediction errors and improving the coefficient of determination $R^2$. We do not optimize models for any particular reaction but aim at predicting the synthesis conditions over a dataset of several thousand synthesis reactions. As demonstrated by the cross-validation and OOS evaluations in Fig. \ref{fig:non-linear-models} and Fig. \ref{fig:oos}, our models achieve $R^2 \sim 0.5-0.6$ (MAE $\sim 140 ^\circ C$) for heating temperature predictions and $R^2 \sim 0.3$ (MAE $\sim 0.3 \log_{10}(h^{-1})$) for heating time predictions. When evaluating these $R^2$ values, it is important to consider that heating temperature and time do not have a single value for a synthesis reaction, as compounds can often be synthesized over a broad range of time and temperature. As such, our models may be more successful at predicting reaction conditions that successfully created the target, as surmised from the $R^2$ scores.

Based on the ranking of DI values in Fig. \ref{fig:di}, the deciding factors for the synthesis conditions can be organized into a two-level hierarchy. Synthesis temperature prediction is dominated by precursor properties, which we speculate are proxies for reactivity stemming from the mobility of ions, with additional corrections learned for different chemistries. Synthesis time prediction is dominated by experiment-adjacent features that are linked to experimental setups/intentions, also with corrections according to chemistry. The features used in this work to account for reaction thermodynamics were inspired by recent efforts to understand phase evolution during synthesis \cite{todd2021selectivity,wustrow2021lowering,miura2021observing,bianchini2020interplay,miura2020selective}. These features involve decomposing overall synthesis reactions into a sequence of phase evolution reactions between pairs of compounds and quantifying the grand potential thermodynamic driving force for these phase evolution reactions. This approach has been proved especially useful for understanding phase evolution pathways observed in \textit{in-situ} experiments. However, in this work, they are shown to provide little predictive power of synthesis conditions and even cause the models to generalize poorly on OOS datasets (as demonstrated in Fig. \ref{fig:oos}). This discrepancy will be discussed in more detail in the subsequent sections.

\subsection{Synthesis adjacent information}

\begin{figure}[t]
  \includegraphics[width=0.55\textwidth]{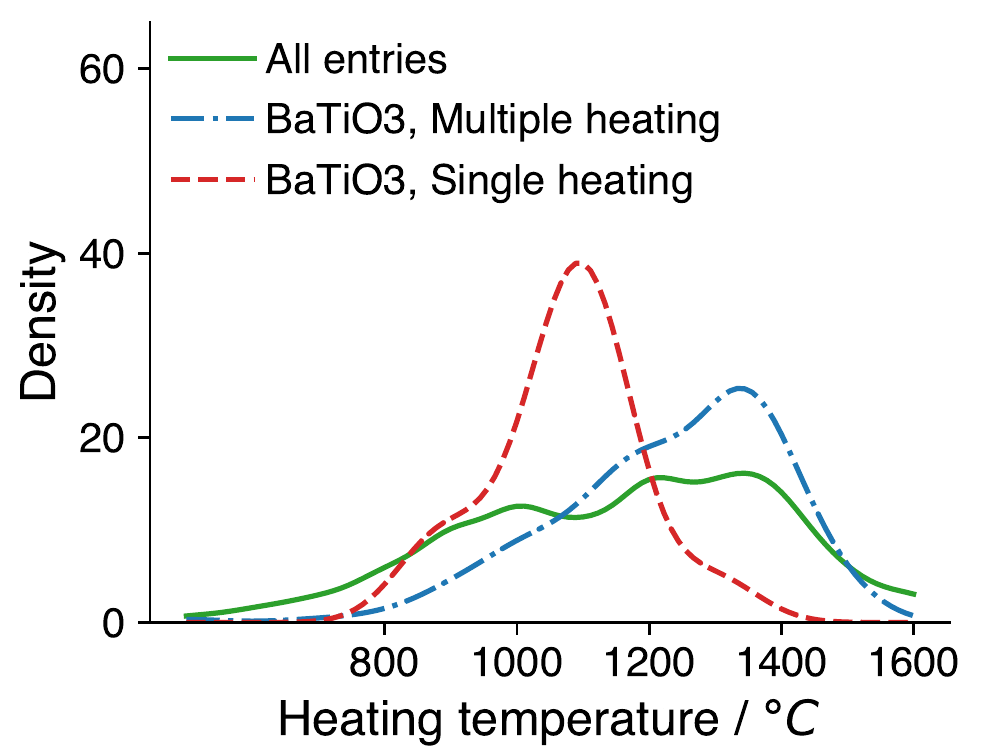}
  \caption{The curves are the distribution of heating temperatures for each group of reactions in the training dataset. The dashed/dotted lines show temperature distributions for the reaction $\ce{TiO2} + \ce{BaCO3} \to \ce{BaTiO3} + \ce{CO2}$ (red dashed line for single-heating reactions and blue dotted line for multiple-heating reactions). Green solid line shows the temperature distribution for the entire dataset.}
  \label{fig:BaCO3-example}
\end{figure}

% Up to this point, our analysis has focused heavily on broad statistical analysis but not on individual systems. 
% {\color{brown}{To Gerd: This section was intended to explain how ML learns synthesis. But I start to feel like its disconnected from the rest of the paper. Therefore, I'd like to remove this section completely. What do you think?}}

We use the particular synthesis of \ce{BaTiO3} from \ce{BaCO3} and \ce{TiO2} precursors to demonstrate how ML models combine synthesis adjacent information with the other regressors. \ce{BaTiO3} is a popular compound with many applications in materials science and appears more than 100 times as the synthesis target in the TMR dataset. A variety of synthesis temperatures have been reported for \ce{BaTiO3} in the literature. For example, \ce{BaTiO3} has been synthesized at $1000 \mathrm{^\circ C}$ \cite{singh2017intense}, $1100 \mathrm{^\circ C}$ \cite{munakata2018effect}, $1200 \mathrm{^\circ C}$ \cite{alluri2017worm}, $1300 \mathrm{^\circ C}$ \cite{zheng2013atomistic}, and $1400 \mathrm{^\circ C}$ \cite{zheng2012grain}. Here we focus on the effect of how many heating steps are used in the synthesis of \ce{BaTiO3}. Fig. \ref{fig:BaCO3-example} shows the distribution of heating temperatures for all the reactions, \ce{BaTiO3} with a single heating step, and \ce{BaTiO3} with multiple heating steps in the training dataset. It is clear that the reported heating temperatures with a single heating step have a lower center around $1100 \mathrm{^\circ C}$ (for example, see ref. \citenum{munakata2018effect}), while the entries with multiple heating steps have a higher center around $1300-1400 \mathrm{^\circ C}$ (for example, see ref. \citenum{zheng2012grain}).

As a result, adding the target composition and experiment-adjacent features allows ML models to identify different groups of data as in Fig. \ref{fig:BaCO3-example} and optimize the predicted heating temperature within each group. For example, if 0 means single heating and 1 means multiple heating, then the ML model should have a coefficient for the feature of ``is multiple heating'' of about $250 \mathrm{^\circ C}$, roughly equal to the difference between the centers of the two temperatures distributions in Fig. \ref{fig:BaCO3-example}.

\subsection{Connection to Tamman's rule}

Our finding that the average precursor melting points are the most predictive feature for heating temperatures is reminiscent of Tamman's rule \cite{tammann1932lehrbuch,merkle2005tammann}. Tamman’s rule can be formulated as predicting that the synthesis temperature of metal alloys should be more than $\frac{1}{3}$ (for example, $\frac{1}{2}-\frac{2}{3}$) of precursor melting points. This rule is derived from the observation that atomic diffusion quickly ceases below $\frac{1}{3}$ of melting temperatures \bibnote{The original German text by Tamman is ``Die Zahl der Platzwechsel in der Zeiteinheit nimmt vom Schmelzpunkt an mit sinkender Temperatur schnell ab und wird bei Metallen bei Metallen bei 1/3 der absoluten Schmelztemperatur unmerklich." which translates to ``The number of changes of place in the unit of time decreases rapidly from the melting point with falling temperature and becomes imperceptible for metals at 1/3 of the absolute melting temperature."}. Tamman's empirical rule was never formally defined. It is also questionable whether the rule is applicable to the synthesis of ionic compounds in addition to intermetallics. Nevertheless, variants of Tamman's rule are still used to help determine solid-state synthesis conditions. For example, \citeauthor{becker2016first} used $\frac{2}{3}$ of the most ``volatile" compound \cite{becker2016first} ; other values, such as $\frac{1}{2}$, have also been used \cite{merkle2005tammann}.

Our ML framework allows us to formally model and test Tamman's rule within a statistical approach. We start with Tamman's original formulation and fit a linear model without an intercept term: $$T_{\mathrm{Tamman}} = \alpha (\min T_{\mathrm{melt}}) + \varepsilon,$$ where $T_{\mathrm{Tamman}}$ is the predicted heating temperature, $(\min{T_{\mathrm{melt}}})$ is the minimum of precursor melting points, $\alpha$ is a parameter to be learned, and $\varepsilon$ is an error term. Both the prediction and the melting points are presented in degrees Kelvin. The fit linear model finds $\alpha = 1.2$ when trained on carbonate reactions and $\alpha=0.8$ when trained on non-carbonate reactions. These $\alpha$ values are larger than the commonly used values for Tamman's rule, such as $1/2$ and $2/3$, suggesting the required temperatures for atoms to diffuse significantly in ionic compounds are higher than in intermetallics, or that for ionic compounds Tamman's rule is a surrogate for another property than diffusion.

The above linear model is not the model with highest predictive power ($R^2$ values). As shown in Fig. \ref{fig:di}, using average precursors melting points (instead of minimum precursor melting points) yields the highest prediction performance. Therefore, we update Tamman's rule to give the optimal synthesis temperature $T_{\mathrm{Tamman}}$ as proportional to the average of precursor melting points $(\mathrm{avg}\:{T_{\mathrm{melt}}})$ plus a constant. Mathematically, the predictor is defined as:
$$T_{\mathrm{Tamman}} = \alpha\: (\mathrm{avg}\:{T_{\mathrm{melt}}}) + \beta + \varepsilon,$$
where $\alpha$, $\beta$ are parameters to be learned and $\varepsilon$ is an error term.

\begin{figure}[ht]
  \begin{subfigure}{0.45\textwidth}
    \centering
    \includegraphics[width=0.95\textwidth]{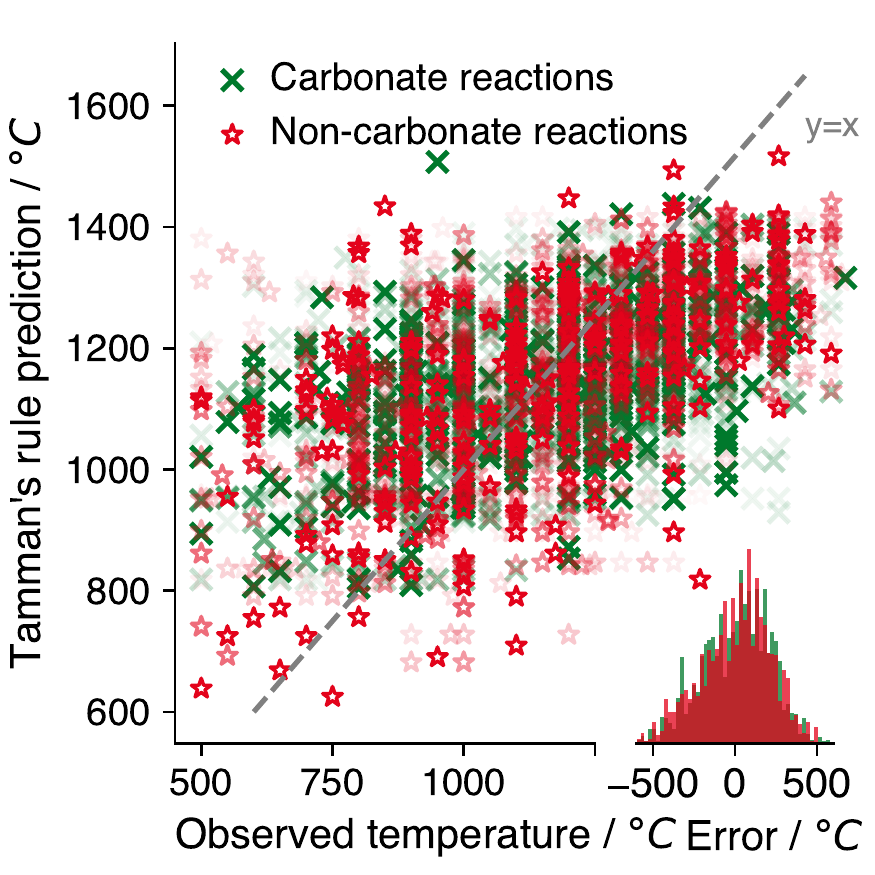}
    \caption{}
    \label{fig:tamman-fit-plot}
  \end{subfigure}
  \begin{subfigure}{0.51\textwidth}
    
    \fontfamily{phv}\selectfont
    \centering
    \begin{tabular}{ccc}
    \hline
    \multicolumn{3}{c}{Fit formula: $T_{\mathrm{Tamman}} = \alpha \:\mathrm{avg}\:{T_{\mathrm{melt}}} + \beta + \varepsilon$}\\
    \hline
    Data subset             & Carbonate                             & Non-carbonate                                   \\
    \hline
    $\hat{\mathbf{\beta}}$  & \textbf{669.3 $\mathrm{^\circ C}$}    & \textbf{569.9 $\mathrm{^\circ C}$}              \\
    $\hat{\mathbf{\alpha}}$ & \textbf{0.346}                        & \textbf{0.329}                                  \\
    $R^2$                   & 0.226                                 & 0.327                                           \\
    \#Obs                   & 3182                                  & 3143                                            \\
    F-statistic             & 931.7                                 & 1526.0                                          \\
    P-value                 & $10^{-179}$                           & $ 10^{-272}$                                    \\
    \hline
    \end{tabular}
    \caption{}
    \label{fig:tamman-fit-table}
  \end{subfigure}
  \caption{Fitting result of Tamman’s rule, i.e., synthesis temperature is proportional to average precursor melting points. \textbf{(a)} Scatter plot of the reported v.s. predicted synthesis temperatures and histogram of prediction error. Opacity indicates data point weights. \textbf{(b)} Regression parameters and F-test for model significance. A very small p-value indicates that it is extremely unlikely the result is due to random noise.}
  \label{fig:tamman-fit}
\end{figure}

As demonstrated in Fig. \ref{fig:tamman-fit}, fitting a linear model reveals a slope of $\sim 1/3$. Since we used the average of precursor melting points, the predicted heating temperatures should be generally larger than $\frac{1}{3}$ of the minimal precursor melting point, agreeing with Tamman's original observation \cite{tammann1932lehrbuch}. The predicted versus reported heating temperatures and the histogram of prediction errors are shown in Fig. \ref{fig:tamman-fit} (a). The parameters of the fitted linear model are shown in Fig. \ref{fig:tamman-fit} (b). The large F-statistic values and very small p-values show strong statistical significance of the model although this is contrasted by the low coefficient of determination ($R^2 \sim 0.2-0.3$). Tamman's rule is not a perfect predictor and has larger prediction errors at low temperatures. However, it contributes more than $\frac{1}{3}$ of the maximal predictive power developed in this work.

\subsection{Roles of phase evolution reaction analysis in synthesis condition prediction}

Predicting heating temperature is of major scientific interest. In solid-state synthesis, the final products are more sensitive to the heating temperature than time, since insufficiently low or high temperatures lead to incomplete reactions, impurities, or the complete absence of a desired target phase. 
% Actually I don't think the following sentence is true. Another impurity may start to form slowly as we heat for longer time.
% On the other hand, setting a much longer heating time usually would not cause too much trouble, as the target phase would have already formed and stabilized long before cooling down. 
Thus, heating temperatures are more carefully optimized than heating times, which are often chosen for convenience (e.g., to run overnight). There have been many successful examples where solid-state synthesis pathways are rationalized using the  thermodynamics of reactions occurring during heating. For example, thermodynamic driving forces have been used to understand and control phase evolution pathways in \ce{Y-Mn-O} oxides \cite{todd2021selectivity,wustrow2021lowering}, \ce{Y-Ba-Cu-O} superconductors \cite{miura2021observing}, \ce{Na-Co-O} layered oxides \cite{bianchini2020interplay}, and \ce{MgCr2S4} thiospinel compounds \cite{miura2020selective}. Inspired by this work, we computed features as numerical transformations of the thermodynamic driving forces obtained by decomposing synthesis into multi-step phase evolution paths. Contrary to the success in reconciling experimental observations in these systems, these features are shown to provide no observable predictive powers for general synthesis condition predictions in this work (as shown in Fig. \ref{fig:di} and Fig. \ref{fig:oos}).

A low contribution of predictive power does not necessarily negate the effectiveness of phase evolution reaction analysis for understanding solid-state synthesis. It simply suggests that the features developed in this work are not correlated with the synthesis time and temperature over the diverse datasets evaluated in this work. We hypothesize this arises for a few reasons. First, the scale of reaction driving force may dictate the decision boundary of synthesizable/non-synthesizable conditions (e.g., synthesis should not occur at temperatures where the target phase is unstable with respect to decomposition). However, the dataset used here only contains positive experimental results, so the thermodynamic stability of the target under the chosen synthesis conditions is likely already achieved for all data points. Indeed, in the rationalization of \textit{in-situ} synthesis, characterization has been used more to explain the phases observed along the reaction path rather than the specific conditions \cite{bianchini2020interplay,miura2021observing,todd2021selectivity}. Second, once we are in the region of synthesizable conditions, reaction driving force might become insufficient in determining synthesis conditions that lead to ``\textit{fast}" reactions. Since a typical lab synthesis needs to be completed in a reasonable period of time, experimenters may decide to raise heating temperatures to facilitate better reaction rates. Indeed, if we calculate the temperature $T_{equilibrium}$ at which the reaction driving force is zero for the overall synthesis reaction (using the grand potential, $\Delta \Phi_{rxn} = 0$) for all the reactions, we found that this theoretical lower bound of heating temperatures $T_{equilibrium}$ is generally much lower than reported experimental $T_{exp}$. This suggests experimenters actively use $T_{exp} \gg T_{equilibrium}$ to achieve better kinetics. Unfortunately, reaction driving force analysis do not directly provide kinetic information, which is also chemistry-specific. On the other hand, precursor melting points and formation energies ($\Delta G_f^{300K}$, $\Delta H_f^{300K}$) may be correlated to ion transport kinetics as they are indicative of the relative strength of bonds in the solid precursors. This may explain why precursor material properties are the top predictive features for heating temperatures.

Previously, we demonstrated that precursor melting points (akin to Tamman's rule) provide the most predictive power for heating temperatures if only one feature is allowed (see IDI values in Fig. \ref{fig:di}). We note here that the effectiveness of Tamman's rule may also be due to the aforementioned selection bias \cite{berger2003randomization} towards fast solid-state syntheses (as well as community knowledge of Tamman's rule). This selection bias is inherent in the synthesis dataset used in this work as the literature only reports ``fast" and successful solid-state reactions. We note that some recent investigations of solid-state synthesis mechanisms \cite{miura2021observing,cosby2020salt} have put more emphasis on modeling reaction speeds. In addition, with the recent developments of autonomous synthesis robots \cite{szymanski2021toward,kimmig2021digital,chen2018exploring,ortiz2019towards}, data on synthesizability and reaction speeds could be collected at the same time with a much higher throughput. Such data will be valuable for decorrelating selection bias and developing broadly applicable synthesis condition predictors.
% Thus, predictions from these models may not align well with the distribution of reported synthesis conditions in the literature. 

\subsection{Challenges of predicting synthesis conditions using text-mined data}

The performance of the ML models in this work is reasonable, but there is still large room for improvements to expand applicability in practical synthesis design efforts. As potential improvements in the future, we summarize a few important aspects for increasing model performance:

\paragraph{Better synthesis features.} Features are limiting factors in creating ML models with high predictive power. This work used 133 features spanning four categories: precursor material properties, target material compositions, reaction thermodynamics, and experiment-adjacent features. Besides these features, one set of useful features may be further factors that indicate the intention of syntheses. For example, the application for which the target compound is created (battery materials vs. thermoelectric materials), desired microstructure of the target materials morphology (single-crystal or spin-coated materials), etc., may all play a roll in the determination of synthesis conditions. These features are expressed in papers in more subtle ways and could be potentially text-mined using advanced NLP techniques in the future \cite{weston2019named,walker2021impact}.

\paragraph{Improved NLP data collection.} Due to the probabilistic nature of the text-mining pipeline that extracted the datasets in this work, errors in the training data are inevitable \cite{kononova2019text}. Manual inspection reveals that 5\% of heating temperatures and  16\% of heating times were incorrectly extracted. Improved text-mining algorithms can thus improve data quality and increase ML model performance.

\paragraph{Modeling non-uniqueness.} In this work, we modeled synthesis condition predictions as point value regression problems. However, this may be sub-optimal, as the conditions where a given synthesis can proceed are non-unique and often span a range of values. Consequently, there is not a unique ground truth of optimal synthesis conditions, which brings irreducible error to ML models. The issue of non-uniqueness is even more problematic for heating time prediction. If the synthesis finishes within $t_0$, then any heating time $t>t_0$ will yield the desired compound, if it is thermodynamically stable at the synthesis conditions and no selective evaporation of elements occurs. As a result, heating time is seldom optimized but based heavily on furnace heating schedule, lab shifts, etc. Indeed, in Fig. \ref{fig:non-linear-models}, our ML models have larger error for predicting heating time than heating temperature. 

Modeling synthesis conditions as distributions, e.g., generalized linear models \cite{nelder1972generalized}, could in principle solve this issue. Note that sufficient training samples must be collected to get accurate condition distribution estimations (as well as uncertainties). Ideally, there would be several conditions sampled for each target that was synthesized in the dataset. However, in the TMR dataset, even when expanding the search to chemical systems (any targets having the same set of elements), more than 60\% contain less than 5 reported syntheses.
Furthermore, the distribution learned from the TMR dataset may be biased by external factors. For example, for popular Li-ion cathode/anode materials in our dataset, the distribution of different synthesis conditions may be correlated with the desired microstructure for a particular electrochemical performance. Decorrelating these factors requires mining of other features/properties beyond the synthesis reactions themselves.

\paragraph{Negative samples.} Negative experimental results are rarely reported in papers. Nevertheless, from a ML point of view, negative data are extremely useful for learning the exact decision boundaries of synthesis conditions. Besides, negative data can be used in other classification tasks, such as predicting the type of synthesis techniques, heating atmospheres, etc.
\\
\\
Finally, we note that the models in this work focused primarily on oxides, which make up a substantial fraction of inorganic compounds but not all\cite{jain2013commentary}. Transferring predictive models trained on oxides to other chemistries is challenging due to significant concept drift. For example, the bonding of other types of compounds, such as non-oxide chalcogenides and intermetallics, is fundamentally different than that of oxides, leading to different self-diffusion and interdiffusion rates. This difference modifies the distributions of feature values significantly (e.g., melting points are systematically lower for metal precursors compared to oxides).
If simply applied to other chemistries without any re-training, the parameters fit for oxide compounds would systematically mis-predict the synthesis conditions.
However, if sufficient data becomes available for desired non-oxide materials classes of interest, the methods used in this work would be useful for training and interpreting these new models.

\section{Conclusion}

In this work, we have developed an interpretable ML method for predicting solid-state synthesis heating temperatures and times on over 6300 reactions synthesis reactions, which are from a larger (over 30,000) synthesis dataset text-mined from scientific literature \cite{kononova2019text}. The goodness-of-fit values are $R^2\sim 0.5-0.6$ for temperature prediction and $R^2\sim 0.3$ for time prediction. Though interpretation of such $R^2$ values has to consider the fact that there is no single exact time or temperature for a typical synthesis. For heating temperature prediction, which is an important parameter for solid-state synthesis, the prediction MAE of our model is $\sim 140 \mathrm{^\circ C}$, comparable to a similar study using generative conditional variational autoencoder (CVAE) \cite{karpovich2021inorganic}. Heating time prediction has a MAE of $\sim 0.3 \log_{10}(h^{-1})$, which translates to a prediction range $[0.5t, 2t]$ if the predicted time is $t$. The expected prediction errors can be estimated from Fig. \ref{fig:violin-error}.

Analysis of the ML models reveals that melting points and formation energies of precursors are good predictors for heating temperatures, which led us to extend Tamman's rule from intermetallics to oxide compounds for predicting heating temperatures as linearly proportional to the average precursor melting points. One may use this extended Tamman's rule to set quick, yet reasonable, initial heating temperatures for new solid-state reactions. The maps of compositional effects (Fig. \ref{fig:elemental-effects}) can be further used as guides to choose synthesis conditions with better accuracy given the chemistries of interest. Our model was trained and validated on a diverse set of materials and thus has broad applicability. Moreover, the ML methodologies developed in this work can be applied for learning synthesis conditions on other large synthesis datasets, such as solution-based synthesis of inorganic compounds and nanoparticles \cite{wang2021dataset,Cruse2021}, or even other tasks where strong model interpretability is preferred.

\section{Methods}

% \paragraph{Discussion of predictive features and trianing data}
% Much similar to a apprentice practice, ML methods acts extracts .
% Characterization of predictive features for synthesis conditions
% We divide potentially interesting features into “empirical” and “fundamental” sets. 
% Empirical features:
% “Rules-of-thumb” values determined by unique chemistries.
% Synthesis-adjacent operations such as ball-milling, regrinding.
% Fundamental features:
% Thermodynamics, hypothesized to dominate solid-state synthesis due to high-T.
% Kinetics of ions, which may affect the diffusion process and therefore reactivity.

\subsection{Curation of synthesis training data}

We used the dataset of text-mined synthesis recipes that consists of 30,004 solid-state synthesis records \cite{kononova2019text} to generate the TMR dataset. We took the synthesis conditions of the last heating step in the experimental procedures as the target of prediction. The synthesis heating temperatures were predicted in degrees Celsius. The reported heating times were transformed to $\log_{10}(1/t)$ which is not only a better variable for measuring reaction speed, but also shows smaller skewness and long tailedness, which is better predicted by statistical ML models \cite{friedman2017elements}. Note that the TMR dataset is extracted using ML models and contains errors in synthesis conditions. Based on manual inspection, about 5\% of the heating temperatures and 16\% of the heating times were incorrectly extracted.

To pre-process the dataset, we first removed all entries with no extracted synthesis heating temperatures and times. To obtain thermodynamic data for all targets, we utilized the Materials Project (MP) database \cite{jain2013commentary}. For targets that appear as entries in MP, we simply used the reported thermodynamic information. For targets without a direct match to an MP entry, we performed interpolation by representing them using linear combinations of the most similar entries in the MP as measured by the difference in composition (see Supplementary Materials for calculation details). The 0 K thermodynamic data was then transformed to finite-temperature Gibbs free energies of formation using the previously developed method \cite{bartel2018physical}.
% Just trying to make them appear in SI.
\nocite{bartel2020critical,bartel2019role}

Using the finite-temperature $\Delta G_f(T)$ predictions and thermodynamic properties of gases, we computed reaction driving forces, i.e., the grand potential change for the synthesis reactions, $\Delta \Phi_{rxn}$, by assuming the system is open to atmospheric partial pressures of \ce{O2} and \ce{CO2} \cite{bartel2022review}. The reactions were then decomposed into phase evolution steps by selecting pairs of reactants with the largest grand potential change in each step. Details of the thermodynamic quantity calculation and phase evolution construction can be found in the Supplementary Material and reproduced using the provided codes. 

We removed the reactions that cannot be handled by the above thermodynamic calculations (e.g., missing relevant MP entries or containing gases other than \ce{O2} and \ce{CO2}), leading to 7,562 remaining reactions. Due to the release of \ce{CO2} gases in carbonate precursor materials, the reaction driving forces have systematically shifted distributions for reactions with and without carbonate precursors. Grouping the dataset into carbonate and non-carbonate reactions thus fits two sets of coefficients that account for this shift and improves the overall performance. Therefore, in our analysis, we split the dataset into carbonate reactions and non-carbonate reactions.

The original Pearson's Crystal Data (PCD) collection is semi-structured containing chemical formulas of input/output materials and a natural language description of the synthesis procedure. We used the same approach as in the generation of the TMR dataset to balance synthesis reactions and calculate phase evolution reaction thermodynamic driving forces. The synthesis procedure description text is used to text-mine synthesis operations that contain synthesis condition values. To make the PCD dataset have similar chemistry distribution as the TMR dataset, we only kept oxide syntheses as the TMR dataset is dominated by oxide syntheses. We also ensured there are no duplicates by removing any entries in the PCD dataset that are also in the TMR dataset by matching their article DOIs.% using the Crossref query service.

\subsection{Features for synthesis prediction}

For each reaction in the curated training data, we computed four types of synthesis features (133 features in total). 

\paragraph{Precursor compound properties.} The first type of features (12 in total) are the average/ minimum/ maximum/ difference of melting points, standard enthalpy of formation $\Delta H_f^{300K}$, standard Gibbs free energy of formation $\Delta G_f^{300K}$ of precursors. The melting points were retrieved from the NIST Chemistry WebBook \bibnote{NIST Chemistry WebBook. \href{https://webbook.nist.gov/chemistry/}{https://webbook.nist.gov/chemistry/} (accessed 2022-07-16).} and PubChem databases \cite{kim2021pubchem}, while the thermodynamic properties were retrieved from the FREED database \bibnote{FREED–Thermodynamic Database. \href{https://www.thermart.net/freed-thermodynamic-database/}{https://www.thermart.net/freed-thermodynamic-database/} (accessed 2022-07-16)}, an electronic compilation of the U.S. Bureau of Mines (USBM) thermodynamic data obtained with experiment. 

\paragraph{Target compound compositional features.} The second type of features are 74 indicator variables representing the presence (1) or absence (0) of different chemical elements in the target compound. We did not use more differentiating features such as the fractional compositions of each element because more than 60\% of the chemical systems in the TMR dataset have less than 5 samples, and more differentiating features make ML models prone to overfitting. Note that this may not be true if training data were to become relatively abundant for each chemical system, in which case numerical encoding of the compositions may be a better approach.

\paragraph{Reaction thermodynamics features.} We used 33 thermodynamic features, including the total reaction driving force $\Delta \Phi_{rxn}$, first and last pairwise reaction driving force $\Delta \Phi_{rxn,1}$, $\Delta \Phi_{rxn,-1}$, and the ratio between first/last pairwise reaction driving force and the total reaction driving force, evaluated at different temperatures $T=800, 900, 1000, 1100, 1200, \text{and}$ $ 1300\: ^\circ C$. We also calculated the slope of $\Delta \Phi_{rxn}, \Delta \Phi_{rxn,1}, \text{and}\: \Delta \Phi_{rxn,-1}$ by assuming they are linear with respect to temperature and used the slopes as additional features.

\paragraph{Experiment-adjacent features.} The fourth type of features are 14 experiment-adjacent features, i.e., indicator variables representing whether certain devices (zirconia balls for ball-milling), experimental procedures (sintering, ball-milling, multiple heating steps, homogenization, repeated grinding, diameter measurement, polycrystalline preparation), and additives (binder materials, distilled water and other liquid additives, phosphors, polyvinyl alcohol)  were used in the synthesis.

Since we used WLS which is sensitive to outliers, we performed outlier detection algorithms on the feature values and removed around 10\% of reactions. The final training data consists of two datasets totaling 6325 reactions. The subset of carbonate reactions consists of 3,182 reactions. The subset of non-carbonate reactions consists of 3,143 reactions.

\subsection{Training and evaluation of ML models}

We used linear and non-linear regressors to train the ML models. For linear models, we used WLS, a weighted version of ordinary least squares in Python packages \textit{scikit-learn} \cite{scikit-learn} and \textit{statsmodels} \cite{seabold2010statsmodels}. For non-linear models, we used the XGBoost package \cite{chen2016xgboost} and trained GBRT models. To evaluate model goodness-of-fit, we used the coefficient of determination, R-squared (or $R^2$). For non-linear regressors and out-of-sample evaluations, $R^2$ is poorly defined and Efron's extended version \cite{efron1978regression} of Pseudo-$R^2$ was used. Pseudo-$R^2$ is calculated as $1 - (\mathrm{Mean\: Square\: Error}/\mathrm{Variance\: of\: data})$ and directly comparable to $R^2$ values.

We implemented DI analysis, a model-agnostic method that calculates the average increase of model $R^2$ to rank features according to their contribution of predictive powers. Three types of DI values, APDI values, IDI values, and IADI values were computed according to \citeauthor{azen2003dominance} \cite{azen2003dominance}. However, to compute the exact APDI values for all the 133 features, we needed to train $2^{133}$ (sub-)models, which is a computationally prohibitive task. Instead, we estimated APDI values $\overline{\Delta (R^2)}$ by randomly sampling 200 submodels for each feature. All the features were ranked according to the sum of APDI, IDI, and IADI values. This ranking measures the relative predictive powers of the features and was used to sort all features in to an ordered list, as in Fig. \ref{fig:di}. 

We next used the ranking of predictive power to perform forward feature selection for the ML models. Specifically, we started with a linear model with no features but the intercept term. Features were sequentially added into the linear model according to the ranking of predictive power. In this process, we calculated the BIC value of the linear models and removed any feature that would increase the BIC value (an indicator of overfitting). The final list of features were then used in training the models in Fig. \ref{fig:non-linear-models} and Fig. \ref{fig:oos}.

We performed LOOCV to cross-validate regressors and detect overfitting. To test model generalizability, we applied out-of-sample prediction by evaluating model performances on another synthesis conditions dataset compiled from the PCD dataset \cite{villars2021}.

\subsection{Code availability}

All codes and data needed to reproduce the results can be found at this repository: \url{https://github.com/CederGroupHub/s4}.

\subsection{Supporting Information}

\begin{itemize}
    \item Further details about the calculation of synthesis predictive features and the construction of machine-learning models
\end{itemize}

% \subsection{References}

% superscript citation style \cite{Mena2000,Abernethy2003}.
% cite authors: \citeauthor{Abernethy2003} and cite year \citeyear{Cotton1999}, or as given by
% Ref.~\citenum{Mena2000}.

% Multiple citations to be combined into a list can be given as
% a single citation.  This uses the \textsf{mciteplus} package
% \cite{Johnson1972,*Arduengo1992,*Eisenstein2005,*Arduengo1994}.
% Citations other than the first of the list should be indicated
% with a star. If the \textsf{mciteplus} package is not installed,
% the standard bibliography tools will still work but starred
% references will be ignored. Individual references can be referred
% to using \texttt{\textbackslash mciteSubRef}:
% ``ref.~\mciteSubRef{Eisenstein2005}''.

% The class also handles notes to be added to the bibliography.  These
% should be given in place in the document \bibnote{This is a note.
% The text will be moved the the references section.  The title of the
% section will change to ``Notes and References''.}.  As with
% citations, the text should be placed before punctuation.  A note is
% also generated if a citation has an optional note.  This assumes that
% the whole work has already been cited: odd numbering will result if
% this is not the case \cite[p.~1]{Cotton1999}.

%%%%%%%%%%%%%%%%%%%%%%%%%%%%%%%%%%%%%%%%%%%%%%%%%%%%%%%%%%%%%%%%%%%%%
%% The "Acknowledgement" section can be given in all manuscript
%% classes.  This should be given within the "acknowledgement"
%% environment, which will make the correct section or running title.
%%%%%%%%%%%%%%%%%%%%%%%%%%%%%%%%%%%%%%%%%%%%%%%%%%%%%%%%%%%%%%%%%%%%%
\begin{acknowledgement}

The authors thank Dr. Olga Kononova for useful discussions on the solid-state synthesis dataset.
This work is supported by the National Science Foundation under DMREF Grant No. DMR-1922372.
Work by A.D. and A.J. was funded by the U.S. Department of Energy, Office of Science, Office of Basic Energy Sciences, Materials Sciences and Engineering Division under contract no. DE-AC02-05-CH11231 (D2S2 program KCD2S2).
This work used the Extreme Science and Engineering Discovery Environment (XSEDE) GPU resources, specifically the Bridges-2 supercomputer at the Pittsburgh Supercomputing Center, through allocation TG-DMR970008S. This work also used computational resources sponsored by the Department of Energy’s Office of Energy Efficiency and Renewable Energy, located at NREL.

\end{acknowledgement}

%%%%%%%%%%%%%%%%%%%%%%%%%%%%%%%%%%%%%%%%%%%%%%%%%%%%%%%%%%%%%%%%%%%%%
%% The appropriate \bibliography command should be placed here.
%% Notice that the class file automatically sets \bibliographystyle
%% and also names the section correctly.
%%%%%%%%%%%%%%%%%%%%%%%%%%%%%%%%%%%%%%%%%%%%%%%%%%%%%%%%%%%%%%%%%%%%%
\bibliography{paper}

\providecommand{\latin}[1]{#1}
\makeatletter
\providecommand{\doi}
  {\begingroup\let\do\@makeother\dospecials
  \catcode`\{=1 \catcode`\}=2 \doi@aux}
\providecommand{\doi@aux}[1]{\endgroup\texttt{#1}}
\makeatother
\providecommand*\mcitethebibliography{\thebibliography}
\csname @ifundefined\endcsname{endmcitethebibliography}
  {\let\endmcitethebibliography\endthebibliography}{}
\begin{mcitethebibliography}{69}
\providecommand*\natexlab[1]{#1}
\providecommand*\mciteSetBstSublistMode[1]{}
\providecommand*\mciteSetBstMaxWidthForm[2]{}
\providecommand*\mciteBstWouldAddEndPuncttrue
  {\def\EndOfBibitem{\unskip.}}
\providecommand*\mciteBstWouldAddEndPunctfalse
  {\let\EndOfBibitem\relax}
\providecommand*\mciteSetBstMidEndSepPunct[3]{}
\providecommand*\mciteSetBstSublistLabelBeginEnd[3]{}
\providecommand*\EndOfBibitem{}
\mciteSetBstSublistMode{f}
\mciteSetBstMaxWidthForm{subitem}{(\alph{mcitesubitemcount})}
\mciteSetBstSublistLabelBeginEnd
  {\mcitemaxwidthsubitemform\space}
  {\relax}
  {\relax}

\bibitem[Kohlmann(2019)]{kohlmann2019looking}
Kohlmann,~H. Looking into the Black Box of Solid-State Synthesis.
  \emph{European Journal of Inorganic Chemistry} \textbf{2019}, \emph{2019},
  4174--4180, DOI: \doi{10.1002/ejic.201900733}\relax
\mciteBstWouldAddEndPuncttrue
\mciteSetBstMidEndSepPunct{\mcitedefaultmidpunct}
{\mcitedefaultendpunct}{\mcitedefaultseppunct}\relax
\EndOfBibitem
\bibitem[Chamorro and McQueen(2018)Chamorro, and McQueen]{chamorro2018progress}
Chamorro,~J.~R.; McQueen,~T.~M. Progress toward solid state synthesis by
  design. \emph{Accounts of chemical research} \textbf{2018}, \emph{51},
  2918--2925, DOI: \doi{10.1021/acs.accounts.8b00382}\relax
\mciteBstWouldAddEndPuncttrue
\mciteSetBstMidEndSepPunct{\mcitedefaultmidpunct}
{\mcitedefaultendpunct}{\mcitedefaultseppunct}\relax
\EndOfBibitem
\bibitem[Shoemaker \latin{et~al.}(2014)Shoemaker, Hu, Chung, Halder, Chupas,
  Soderholm, Mitchell, and Kanatzidis]{shoemaker2014situ}
Shoemaker,~D.~P.; Hu,~Y.-J.; Chung,~D.~Y.; Halder,~G.~J.; Chupas,~P.~J.;
  Soderholm,~L.; Mitchell,~J.; Kanatzidis,~M.~G. In situ studies of a platform
  for metastable inorganic crystal growth and materials discovery.
  \emph{Proceedings of the National Academy of Sciences} \textbf{2014},
  \emph{111}, 10922--10927, DOI: \doi{10.1073/pnas.1406211111}\relax
\mciteBstWouldAddEndPuncttrue
\mciteSetBstMidEndSepPunct{\mcitedefaultmidpunct}
{\mcitedefaultendpunct}{\mcitedefaultseppunct}\relax
\EndOfBibitem
\bibitem[McClain \latin{et~al.}(2021)McClain, Malliakas, Shen, He, Wolverton,
  Gonz{\'a}lez, and Kanatzidis]{mcclain2021mechanistic}
McClain,~R.; Malliakas,~C.~D.; Shen,~J.; He,~J.; Wolverton,~C.;
  Gonz{\'a}lez,~G.~B.; Kanatzidis,~M.~G. Mechanistic insight of KBiQ2 (Q=S,Se)
  using panoramic synthesis towards synthesis-by-design. \emph{Chemical
  Science} \textbf{2021}, \emph{12}, 1378--1391, DOI:
  \doi{10.1039/D0SC04562D}\relax
\mciteBstWouldAddEndPuncttrue
\mciteSetBstMidEndSepPunct{\mcitedefaultmidpunct}
{\mcitedefaultendpunct}{\mcitedefaultseppunct}\relax
\EndOfBibitem
\bibitem[Ito \latin{et~al.}(2021)Ito, Shitara, Wang, Fujii, Yashima, Goto,
  Moriyoshi, Rosero-Navarro, Miura, and Tadanaga]{ito2021kinetically}
Ito,~H.; Shitara,~K.; Wang,~Y.; Fujii,~K.; Yashima,~M.; Goto,~Y.;
  Moriyoshi,~C.; Rosero-Navarro,~N.~C.; Miura,~A.; Tadanaga,~K. Kinetically
  Stabilized Cation Arrangement in Li3YCl6 Superionic Conductor during
  Solid-State Reaction. \emph{Advanced Science} \textbf{2021}, 2101413, DOI:
  \doi{10.1002/advs.202101413}\relax
\mciteBstWouldAddEndPuncttrue
\mciteSetBstMidEndSepPunct{\mcitedefaultmidpunct}
{\mcitedefaultendpunct}{\mcitedefaultseppunct}\relax
\EndOfBibitem
\bibitem[Paradis-Fortin \latin{et~al.}(2020)Paradis-Fortin, Lemoine,
  Prestipino, Kumar, Raveau, Nassif, Cordier, and Guilmeau]{paradis2020time}
Paradis-Fortin,~L.; Lemoine,~P.; Prestipino,~C.; Kumar,~V.~P.; Raveau,~B.;
  Nassif,~V.; Cordier,~S.; Guilmeau,~E. Time-resolved in situ neutron
  diffraction study of Cu22Fe8Ge4S32 germanite: a guide for the synthesis of
  complex chalcogenides. \emph{Chemistry of Materials} \textbf{2020},
  \emph{32}, 8993--9000, DOI: \doi{10.1021/acs.chemmater.0c03219}\relax
\mciteBstWouldAddEndPuncttrue
\mciteSetBstMidEndSepPunct{\mcitedefaultmidpunct}
{\mcitedefaultendpunct}{\mcitedefaultseppunct}\relax
\EndOfBibitem
\bibitem[Bianchini \latin{et~al.}(2020)Bianchini, Wang, Cl{\'e}ment, Ouyang,
  Xiao, Kitchaev, Shi, Zhang, Wang, Kim, \latin{et~al.}
  others]{bianchini2020interplay}
Bianchini,~M.; Wang,~J.; Cl{\'e}ment,~R.~J.; Ouyang,~B.; Xiao,~P.;
  Kitchaev,~D.; Shi,~T.; Zhang,~Y.; Wang,~Y.; Kim,~H., \latin{et~al.}  The
  interplay between thermodynamics and kinetics in the solid-state synthesis of
  layered oxides. \emph{Nature materials} \textbf{2020}, \emph{19}, 1088--1095,
  DOI: \doi{10.1038/s41563-020-0688-6}\relax
\mciteBstWouldAddEndPuncttrue
\mciteSetBstMidEndSepPunct{\mcitedefaultmidpunct}
{\mcitedefaultendpunct}{\mcitedefaultseppunct}\relax
\EndOfBibitem
\bibitem[Miura \latin{et~al.}(2021)Miura, Bartel, Goto, Mizuguchi, Moriyoshi,
  Kuroiwa, Wang, Yaguchi, Shirai, Nagao, \latin{et~al.}
  others]{miura2021observing}
Miura,~A.; Bartel,~C.~J.; Goto,~Y.; Mizuguchi,~Y.; Moriyoshi,~C.; Kuroiwa,~Y.;
  Wang,~Y.; Yaguchi,~T.; Shirai,~M.; Nagao,~M., \latin{et~al.}  Observing and
  Modeling the Sequential Pairwise Reactions that Drive Solid-State Ceramic
  Synthesis. \emph{Advanced Materials} \textbf{2021}, 2100312, DOI:
  \doi{10.1002/adma.202100312}\relax
\mciteBstWouldAddEndPuncttrue
\mciteSetBstMidEndSepPunct{\mcitedefaultmidpunct}
{\mcitedefaultendpunct}{\mcitedefaultseppunct}\relax
\EndOfBibitem
\bibitem[Miura \latin{et~al.}(2020)Miura, Ito, Bartel, Sun, Rosero-Navarro,
  Tadanaga, Nakata, Maeda, and Ceder]{miura2020selective}
Miura,~A.; Ito,~H.; Bartel,~C.~J.; Sun,~W.; Rosero-Navarro,~N.~C.;
  Tadanaga,~K.; Nakata,~H.; Maeda,~K.; Ceder,~G. Selective metathesis synthesis
  of MgCr2S4 by control of thermodynamic driving forces. \emph{Materials
  horizons} \textbf{2020}, \emph{7}, 1310--1316, DOI:
  \doi{10.1039/C9MH01999E}\relax
\mciteBstWouldAddEndPuncttrue
\mciteSetBstMidEndSepPunct{\mcitedefaultmidpunct}
{\mcitedefaultendpunct}{\mcitedefaultseppunct}\relax
\EndOfBibitem
\bibitem[McDermott \latin{et~al.}(2021)McDermott, Dwaraknath, and
  Persson]{mcdermott2021graph}
McDermott,~M.~J.; Dwaraknath,~S.~S.; Persson,~K.~A. A graph-based network for
  predicting chemical reaction pathways in solid-state materials synthesis.
  \emph{Nature communications} \textbf{2021}, \emph{12}, 3097, DOI:
  \doi{10.1038/s41467-021-23339-x}\relax
\mciteBstWouldAddEndPuncttrue
\mciteSetBstMidEndSepPunct{\mcitedefaultmidpunct}
{\mcitedefaultendpunct}{\mcitedefaultseppunct}\relax
\EndOfBibitem
\bibitem[Aykol \latin{et~al.}(2021)Aykol, Montoya, and
  Hummelsh{\o}j]{aykol2021rational}
Aykol,~M.; Montoya,~J.~H.; Hummelsh{\o}j,~J. Rational solid-state synthesis
  routes for inorganic materials. \emph{Journal of the American Chemical
  Society} \textbf{2021}, \emph{143}, 9244--9259, DOI:
  \doi{10.1021/jacs.1c04888}\relax
\mciteBstWouldAddEndPuncttrue
\mciteSetBstMidEndSepPunct{\mcitedefaultmidpunct}
{\mcitedefaultendpunct}{\mcitedefaultseppunct}\relax
\EndOfBibitem
\bibitem[Wustrow \latin{et~al.}(2021)Wustrow, Huang, McDermott, O’Nolan, Liu,
  Tran, McBride, Dwaraknath, Chapman, Billinge, \latin{et~al.}
  others]{wustrow2021lowering}
Wustrow,~A.; Huang,~G.; McDermott,~M.~J.; O’Nolan,~D.; Liu,~C.-H.;
  Tran,~G.~T.; McBride,~B.~C.; Dwaraknath,~S.~S.; Chapman,~K.~W.;
  Billinge,~S.~J., \latin{et~al.}  Lowering Ternary Oxide Synthesis
  Temperatures by Solid-State Cometathesis Reactions. \emph{Chemistry of
  Materials} \textbf{2021}, \emph{33}, 3692--3701, DOI:
  \doi{10.1021/acs.chemmater.1c00700}\relax
\mciteBstWouldAddEndPuncttrue
\mciteSetBstMidEndSepPunct{\mcitedefaultmidpunct}
{\mcitedefaultendpunct}{\mcitedefaultseppunct}\relax
\EndOfBibitem
\bibitem[Kim \latin{et~al.}(2017)Kim, Huang, Tomala, Matthews, Strubell,
  Saunders, McCallum, and Olivetti]{kim2017machine}
Kim,~E.; Huang,~K.; Tomala,~A.; Matthews,~S.; Strubell,~E.; Saunders,~A.;
  McCallum,~A.; Olivetti,~E. Machine-learned and codified synthesis parameters
  of oxide materials. \emph{Scientific data} \textbf{2017}, \emph{4}, 170127,
  DOI: \doi{10.1038/sdata.2017.127}\relax
\mciteBstWouldAddEndPuncttrue
\mciteSetBstMidEndSepPunct{\mcitedefaultmidpunct}
{\mcitedefaultendpunct}{\mcitedefaultseppunct}\relax
\EndOfBibitem
\bibitem[Kim \latin{et~al.}(2017)Kim, Huang, Saunders, McCallum, Ceder, and
  Olivetti]{kim2017materials}
Kim,~E.; Huang,~K.; Saunders,~A.; McCallum,~A.; Ceder,~G.; Olivetti,~E.
  Materials synthesis insights from scientific literature via text extraction
  and machine learning. \emph{Chemistry of Materials} \textbf{2017}, \emph{29},
  9436--9444, DOI: \doi{10.1021/acs.chemmater.7b03500}\relax
\mciteBstWouldAddEndPuncttrue
\mciteSetBstMidEndSepPunct{\mcitedefaultmidpunct}
{\mcitedefaultendpunct}{\mcitedefaultseppunct}\relax
\EndOfBibitem
\bibitem[Kim \latin{et~al.}(2020)Kim, Jensen, van Grootel, Huang, Staib,
  Mysore, Chang, Strubell, McCallum, Jegelka, \latin{et~al.}
  others]{kim2020inorganic}
Kim,~E.; Jensen,~Z.; van Grootel,~A.; Huang,~K.; Staib,~M.; Mysore,~S.;
  Chang,~H.-S.; Strubell,~E.; McCallum,~A.; Jegelka,~S., \latin{et~al.}
  Inorganic materials synthesis planning with literature-trained neural
  networks. \emph{Journal of chemical information and modeling} \textbf{2020},
  \emph{60}, 1194--1201, DOI: \doi{10.1021/acs.jcim.9b00995}\relax
\mciteBstWouldAddEndPuncttrue
\mciteSetBstMidEndSepPunct{\mcitedefaultmidpunct}
{\mcitedefaultendpunct}{\mcitedefaultseppunct}\relax
\EndOfBibitem
\bibitem[Kononova \latin{et~al.}(2019)Kononova, Huo, He, Rong, Botari, Sun,
  Tshitoyan, and Ceder]{kononova2019text}
Kononova,~O.; Huo,~H.; He,~T.; Rong,~Z.; Botari,~T.; Sun,~W.; Tshitoyan,~V.;
  Ceder,~G. Text-mined dataset of inorganic materials synthesis recipes.
  \emph{Scientific data} \textbf{2019}, \emph{6}, 203, DOI:
  \doi{10.1038/s41597-019-0224-1}\relax
\mciteBstWouldAddEndPuncttrue
\mciteSetBstMidEndSepPunct{\mcitedefaultmidpunct}
{\mcitedefaultendpunct}{\mcitedefaultseppunct}\relax
\EndOfBibitem
\bibitem[Vaucher \latin{et~al.}(2020)Vaucher, Zipoli, Geluykens, Nair,
  Schwaller, and Laino]{vaucher2020automated}
Vaucher,~A.~C.; Zipoli,~F.; Geluykens,~J.; Nair,~V.~H.; Schwaller,~P.;
  Laino,~T. Automated extraction of chemical synthesis actions from
  experimental procedures. \emph{Nature communications} \textbf{2020},
  \emph{11}, 3601, DOI: \doi{10.1038/s41467-020-17266-6}\relax
\mciteBstWouldAddEndPuncttrue
\mciteSetBstMidEndSepPunct{\mcitedefaultmidpunct}
{\mcitedefaultendpunct}{\mcitedefaultseppunct}\relax
\EndOfBibitem
\bibitem[Young \latin{et~al.}(2018)Young, Maksov, Ziatdinov, Cao, Burch,
  Balachandran, Li, Somnath, Patton, Kalinin, \latin{et~al.}
  others]{young2018data}
Young,~S.~R.; Maksov,~A.; Ziatdinov,~M.; Cao,~Y.; Burch,~M.; Balachandran,~J.;
  Li,~L.; Somnath,~S.; Patton,~R.~M.; Kalinin,~S.~V., \latin{et~al.}  Data
  mining for better material synthesis: The case of pulsed laser deposition of
  complex oxides. \emph{Journal of Applied Physics} \textbf{2018}, \emph{123},
  115303, DOI: \doi{10.1063/1.5009942}\relax
\mciteBstWouldAddEndPuncttrue
\mciteSetBstMidEndSepPunct{\mcitedefaultmidpunct}
{\mcitedefaultendpunct}{\mcitedefaultseppunct}\relax
\EndOfBibitem
\bibitem[Karpovich \latin{et~al.}()Karpovich, Jensen, Venugopal, and
  Olivetti]{karpovich2021inorganic}
Karpovich,~C.; Jensen,~Z.; Venugopal,~V.; Olivetti,~E. Inorganic Synthesis
  Reaction Condition Prediction with Generative Machine Learning. 2021-12-17.
  \emph{arXiv preprint arXiv:2112.09612} \url{https://arxiv.org/abs/2112.09612}
  (accessed 2022-07-16)\relax
\mciteBstWouldAddEndPuncttrue
\mciteSetBstMidEndSepPunct{\mcitedefaultmidpunct}
{\mcitedefaultendpunct}{\mcitedefaultseppunct}\relax
\EndOfBibitem
\bibitem[Davariashtiyani \latin{et~al.}(2021)Davariashtiyani, Kadkhodaie, and
  Kadkhodaei]{davariashtiyani2021predicting}
Davariashtiyani,~A.; Kadkhodaie,~Z.; Kadkhodaei,~S. Predicting synthesizability
  of crystalline materials via deep learning. \emph{Communications Materials}
  \textbf{2021}, \emph{2}, 115, DOI: \doi{10.1038/s43246-021-00219-x}\relax
\mciteBstWouldAddEndPuncttrue
\mciteSetBstMidEndSepPunct{\mcitedefaultmidpunct}
{\mcitedefaultendpunct}{\mcitedefaultseppunct}\relax
\EndOfBibitem
\bibitem[Sun and Powell-Palm()Sun, and Powell-Palm]{sun2021generalized}
Sun,~W.; Powell-Palm,~M.~J. Generalized Gibbs' Phase Rule. 2021-05-04.
  \emph{arXiv preprint arXiv:2105.01337} \url{https://arxiv.org/abs/2105.01337}
  (accessed 2022-07-16)\relax
\mciteBstWouldAddEndPuncttrue
\mciteSetBstMidEndSepPunct{\mcitedefaultmidpunct}
{\mcitedefaultendpunct}{\mcitedefaultseppunct}\relax
\EndOfBibitem
\bibitem[Sun \latin{et~al.}(2016)Sun, Dacek, Ong, Hautier, Jain, Richards,
  Gamst, Persson, and Ceder]{sun2016thermodynamic}
Sun,~W.; Dacek,~S.~T.; Ong,~S.~P.; Hautier,~G.; Jain,~A.; Richards,~W.~D.;
  Gamst,~A.~C.; Persson,~K.~A.; Ceder,~G. The thermodynamic scale of inorganic
  crystalline metastability. \emph{Science advances} \textbf{2016}, \emph{2},
  e1600225, DOI: \doi{10.1126/sciadv.1600225}\relax
\mciteBstWouldAddEndPuncttrue
\mciteSetBstMidEndSepPunct{\mcitedefaultmidpunct}
{\mcitedefaultendpunct}{\mcitedefaultseppunct}\relax
\EndOfBibitem
\bibitem[Aykol \latin{et~al.}(2018)Aykol, Dwaraknath, Sun, and
  Persson]{aykol2018thermodynamic}
Aykol,~M.; Dwaraknath,~S.~S.; Sun,~W.; Persson,~K.~A. Thermodynamic limit for
  synthesis of metastable inorganic materials. \emph{Science advances}
  \textbf{2018}, \emph{4}, eaaq0148, DOI: \doi{10.1126/sciadv.aaq0148}\relax
\mciteBstWouldAddEndPuncttrue
\mciteSetBstMidEndSepPunct{\mcitedefaultmidpunct}
{\mcitedefaultendpunct}{\mcitedefaultseppunct}\relax
\EndOfBibitem
\bibitem[Dinia \latin{et~al.}(2004)Dinia, V{\'e}nuat, Colis, and
  Pourroy]{dinia2004elaboration}
Dinia,~A.; V{\'e}nuat,~J.; Colis,~S.; Pourroy,~G. Elaboration and
  characterization of the Sr2FeMoO6 double perovskite. \emph{Catalysis today}
  \textbf{2004}, \emph{89}, 297--302, DOI:
  \doi{10.1016/j.cattod.2003.12.019}\relax
\mciteBstWouldAddEndPuncttrue
\mciteSetBstMidEndSepPunct{\mcitedefaultmidpunct}
{\mcitedefaultendpunct}{\mcitedefaultseppunct}\relax
\EndOfBibitem
\bibitem[Shi \latin{et~al.}(2018)Shi, Xiao, Kwon, Sai~Gautam, Chakarawet, Kim,
  Bo, and Ceder]{shi2018shear}
Shi,~T.; Xiao,~P.; Kwon,~D.-H.; Sai~Gautam,~G.; Chakarawet,~K.; Kim,~H.;
  Bo,~S.-H.; Ceder,~G. Shear-assisted formation of cation-disordered rocksalt
  NaMO2 (M=Fe or Mn). \emph{Chemistry of Materials} \textbf{2018}, \emph{30},
  8811--8821, DOI: \doi{10.1021/acs.chemmater.8b03490}\relax
\mciteBstWouldAddEndPuncttrue
\mciteSetBstMidEndSepPunct{\mcitedefaultmidpunct}
{\mcitedefaultendpunct}{\mcitedefaultseppunct}\relax
\EndOfBibitem
\bibitem[Rao and Biswas(2015)Rao, and Biswas]{rao2015essentials}
Rao,~C. N.~R.; Biswas,~K. \emph{Essentials of inorganic materials synthesis};
  John Wiley \& Sons, 2015\relax
\mciteBstWouldAddEndPuncttrue
\mciteSetBstMidEndSepPunct{\mcitedefaultmidpunct}
{\mcitedefaultendpunct}{\mcitedefaultseppunct}\relax
\EndOfBibitem
\bibitem[Yuan \latin{et~al.}(2011)Yuan, Cai, and Shao]{yuan2011different}
Yuan,~T.; Cai,~R.; Shao,~Z. Different effect of the atmospheres on the phase
  formation and performance of Li4Ti5O12 prepared from ball-milling-assisted
  solid-phase reaction with pristine and carbon-precoated TiO2 as starting
  materials. \emph{The Journal of Physical Chemistry C} \textbf{2011},
  \emph{115}, 4943--4952, DOI: \doi{10.1021/jp111353e}\relax
\mciteBstWouldAddEndPuncttrue
\mciteSetBstMidEndSepPunct{\mcitedefaultmidpunct}
{\mcitedefaultendpunct}{\mcitedefaultseppunct}\relax
\EndOfBibitem
\bibitem[Montgomery \latin{et~al.}(2021)Montgomery, Peck, and
  Vining]{montgomery2021introduction}
Montgomery,~D.~C.; Peck,~E.~A.; Vining,~G.~G. \emph{Introduction to linear
  regression analysis}; John Wiley \& Sons, 2021\relax
\mciteBstWouldAddEndPuncttrue
\mciteSetBstMidEndSepPunct{\mcitedefaultmidpunct}
{\mcitedefaultendpunct}{\mcitedefaultseppunct}\relax
\EndOfBibitem
\bibitem[Friedman(2017)]{friedman2017elements}
Friedman,~J.~H. \emph{The elements of statistical learning: Data mining,
  inference, and prediction}; Springer, 2017\relax
\mciteBstWouldAddEndPuncttrue
\mciteSetBstMidEndSepPunct{\mcitedefaultmidpunct}
{\mcitedefaultendpunct}{\mcitedefaultseppunct}\relax
\EndOfBibitem
\bibitem[Kononova \latin{et~al.}(2021)Kononova, He, Huo, Trewartha, Olivetti,
  and Ceder]{kononova2021opportunities}
Kononova,~O.; He,~T.; Huo,~H.; Trewartha,~A.; Olivetti,~E.~A.; Ceder,~G.
  Opportunities and challenges of text mining in materials research.
  \emph{iScience} \textbf{2021}, \emph{24}, DOI:
  \doi{10.1016/j.isci.2021.102155}\relax
\mciteBstWouldAddEndPuncttrue
\mciteSetBstMidEndSepPunct{\mcitedefaultmidpunct}
{\mcitedefaultendpunct}{\mcitedefaultseppunct}\relax
\EndOfBibitem
\bibitem[Azen and Budescu(2003)Azen, and Budescu]{azen2003dominance}
Azen,~R.; Budescu,~D.~V. The dominance analysis approach for comparing
  predictors in multiple regression. \emph{Psychological methods}
  \textbf{2003}, \emph{8}, 129--148, DOI: \doi{10.1037/1082-989x.8.2.129}\relax
\mciteBstWouldAddEndPuncttrue
\mciteSetBstMidEndSepPunct{\mcitedefaultmidpunct}
{\mcitedefaultendpunct}{\mcitedefaultseppunct}\relax
\EndOfBibitem
\bibitem[Villars and Cenzual(2021)Villars, and Cenzual]{villars2021}
Villars,~P.; Cenzual,~K. \emph{Pearson's Crystal Data: Crystal Structure
  Database for Inorganic Compounds (on DVD)}, release 2021/22 ed.; ASM
  International®: Materials Park, Ohio, USA, 2021\relax
\mciteBstWouldAddEndPuncttrue
\mciteSetBstMidEndSepPunct{\mcitedefaultmidpunct}
{\mcitedefaultendpunct}{\mcitedefaultseppunct}\relax
\EndOfBibitem
\bibitem[Faria and Soromenho(2010)Faria, and Soromenho]{faria2010fitting}
Faria,~S.; Soromenho,~G. Fitting mixtures of linear regressions. \emph{Journal
  of Statistical Computation and Simulation} \textbf{2010}, \emph{80},
  201--225, DOI: \doi{10.1080/00949650802590261}\relax
\mciteBstWouldAddEndPuncttrue
\mciteSetBstMidEndSepPunct{\mcitedefaultmidpunct}
{\mcitedefaultendpunct}{\mcitedefaultseppunct}\relax
\EndOfBibitem
\bibitem[Li and Liang(2018)Li, and Liang]{li2018learning}
Li,~Y.; Liang,~Y. Learning mixtures of linear regressions with nearly optimal
  complexity. Conference On Learning Theory. 2018; pp 1125--1144\relax
\mciteBstWouldAddEndPuncttrue
\mciteSetBstMidEndSepPunct{\mcitedefaultmidpunct}
{\mcitedefaultendpunct}{\mcitedefaultseppunct}\relax
\EndOfBibitem
\bibitem[Seabold and Perktold(2010)Seabold, and
  Perktold]{seabold2010statsmodels}
Seabold,~S.; Perktold,~J. {S}tatsmodels: {E}conometric and {S}tatistical
  {M}odeling with {P}ython. {P}roceedings of the 9th {P}ython in {S}cience
  {C}onference. 2010; pp 92 -- 96, DOI:
  \doi{10.25080/Majora-92bf1922-011}\relax
\mciteBstWouldAddEndPuncttrue
\mciteSetBstMidEndSepPunct{\mcitedefaultmidpunct}
{\mcitedefaultendpunct}{\mcitedefaultseppunct}\relax
\EndOfBibitem
\bibitem[Chen and Guestrin(2016)Chen, and Guestrin]{chen2016xgboost}
Chen,~T.; Guestrin,~C. Xgboost: A scalable tree boosting system. Proceedings of
  the 22nd acm sigkdd international conference on knowledge discovery and data
  mining. 2016; pp 785--794, DOI: \doi{10.1145/2939672.2939785}\relax
\mciteBstWouldAddEndPuncttrue
\mciteSetBstMidEndSepPunct{\mcitedefaultmidpunct}
{\mcitedefaultendpunct}{\mcitedefaultseppunct}\relax
\EndOfBibitem
\bibitem[Qui{\~n}onero-Candela \latin{et~al.}(2009)Qui{\~n}onero-Candela,
  Sugiyama, Lawrence, and Schwaighofer]{quinonero2009dataset}
Qui{\~n}onero-Candela,~J.; Sugiyama,~M.; Lawrence,~N.~D.; Schwaighofer,~A.
  \emph{Dataset shift in machine learning}; Mit Press, 2009\relax
\mciteBstWouldAddEndPuncttrue
\mciteSetBstMidEndSepPunct{\mcitedefaultmidpunct}
{\mcitedefaultendpunct}{\mcitedefaultseppunct}\relax
\EndOfBibitem
\bibitem[Todd \latin{et~al.}(2021)Todd, McDermott, Rom, Corrao, Denney,
  Dwaraknath, Khalifah, Persson, and Neilson]{todd2021selectivity}
Todd,~P.~K.; McDermott,~M.~J.; Rom,~C.~L.; Corrao,~A.~A.; Denney,~J.~J.;
  Dwaraknath,~S.~S.; Khalifah,~P.~G.; Persson,~K.~A.; Neilson,~J.~R.
  Selectivity in Yttrium Manganese Oxide Synthesis via Local Chemical
  Potentials in Hyperdimensional Phase Space. \emph{Journal of the American
  Chemical Society} \textbf{2021}, \emph{143}, 15185--15194, DOI:
  \doi{10.1021/jacs.1c06229}\relax
\mciteBstWouldAddEndPuncttrue
\mciteSetBstMidEndSepPunct{\mcitedefaultmidpunct}
{\mcitedefaultendpunct}{\mcitedefaultseppunct}\relax
\EndOfBibitem
\bibitem[Singh \latin{et~al.}(2017)Singh, Kaur, Bose, Shrivastava, Dubey, and
  Parganiha]{singh2017intense}
Singh,~R.; Kaur,~J.; Bose,~P.; Shrivastava,~R.; Dubey,~V.; Parganiha,~Y.
  Intense visible light emission from dysprosium (Dy3+) doped barium titanate
  (BaTiO3) phosphor and its thermoluminescence study. \emph{Journal of
  Materials Science: Materials in Electronics} \textbf{2017}, \emph{28},
  13690--13697, DOI: \doi{10.1007/s10854-017-7212-z}\relax
\mciteBstWouldAddEndPuncttrue
\mciteSetBstMidEndSepPunct{\mcitedefaultmidpunct}
{\mcitedefaultendpunct}{\mcitedefaultseppunct}\relax
\EndOfBibitem
\bibitem[Munakata \latin{et~al.}(2018)Munakata, Yoshino, Nemoto, Abe, and
  Ito]{munakata2018effect}
Munakata,~F.; Yoshino,~K.; Nemoto,~K.; Abe,~S.; Ito,~A. Effect of self-assembly
  material texture and dielectric properties of BaTiO3/poly-l-lactic-acid
  composites. \emph{Materials Letters} \textbf{2018}, \emph{221}, 147--149,
  DOI: \doi{10.1016/j.matlet.2018.03.008}\relax
\mciteBstWouldAddEndPuncttrue
\mciteSetBstMidEndSepPunct{\mcitedefaultmidpunct}
{\mcitedefaultendpunct}{\mcitedefaultseppunct}\relax
\EndOfBibitem
\bibitem[Alluri \latin{et~al.}(2017)Alluri, Selvarajan, Chandrasekhar,
  Saravanakumar, Lee, Jeong, and Kim]{alluri2017worm}
Alluri,~N.~R.; Selvarajan,~S.; Chandrasekhar,~A.; Saravanakumar,~B.;
  Lee,~G.~M.; Jeong,~J.~H.; Kim,~S.-J. Worm structure piezoelectric energy
  harvester using ionotropic gelation of barium titanate-calcium alginate
  composite. \emph{Energy} \textbf{2017}, \emph{118}, 1146--1155, DOI:
  \doi{10.1016/j.energy.2016.10.143}\relax
\mciteBstWouldAddEndPuncttrue
\mciteSetBstMidEndSepPunct{\mcitedefaultmidpunct}
{\mcitedefaultendpunct}{\mcitedefaultseppunct}\relax
\EndOfBibitem
\bibitem[Zheng \latin{et~al.}(2013)Zheng, Ma, Yamamoto, and
  Ikuhara]{zheng2013atomistic}
Zheng,~S.; Ma,~X.; Yamamoto,~T.; Ikuhara,~Y. Atomistic study of abnormal grain
  growth structure in BaTiO3 by transmission electron microscopy and scanning
  transmission electron microscopy. \emph{Acta materialia} \textbf{2013},
  \emph{61}, 2298--2307, DOI: \doi{10.1016/j.actamat.2012.12.046}\relax
\mciteBstWouldAddEndPuncttrue
\mciteSetBstMidEndSepPunct{\mcitedefaultmidpunct}
{\mcitedefaultendpunct}{\mcitedefaultseppunct}\relax
\EndOfBibitem
\bibitem[Zheng \latin{et~al.}(2012)Zheng, Zhang, Tan, and Wang]{zheng2012grain}
Zheng,~P.; Zhang,~J.; Tan,~Y.; Wang,~C. Grain-size effects on dielectric and
  piezoelectric properties of poled BaTiO3 ceramics. \emph{Acta Materialia}
  \textbf{2012}, \emph{60}, 5022--5030, DOI:
  \doi{10.1016/j.actamat.2012.06.015}\relax
\mciteBstWouldAddEndPuncttrue
\mciteSetBstMidEndSepPunct{\mcitedefaultmidpunct}
{\mcitedefaultendpunct}{\mcitedefaultseppunct}\relax
\EndOfBibitem
\bibitem[Tammann(1932)]{tammann1932lehrbuch}
Tammann,~G. \emph{Lehrbuch der Metallographie: Chemie und Physik der Metalle
  und ihrer Legierungen}; Leopold Voss, Leipzig, 1932; p 314\relax
\mciteBstWouldAddEndPuncttrue
\mciteSetBstMidEndSepPunct{\mcitedefaultmidpunct}
{\mcitedefaultendpunct}{\mcitedefaultseppunct}\relax
\EndOfBibitem
\bibitem[Merkle and Maier(2005)Merkle, and Maier]{merkle2005tammann}
Merkle,~R.; Maier,~J. On the tammann--rule. \emph{Zeitschrift f{\"u}r
  anorganische und allgemeine Chemie} \textbf{2005}, \emph{631}, 1163--1166,
  DOI: \doi{10.1002/zaac.200400540}\relax
\mciteBstWouldAddEndPuncttrue
\mciteSetBstMidEndSepPunct{\mcitedefaultmidpunct}
{\mcitedefaultendpunct}{\mcitedefaultseppunct}\relax
\EndOfBibitem
\bibitem[Not()]{Note-1}
The original German text by Tamman is ``Die Zahl der Platzwechsel in der
  Zeiteinheit nimmt vom Schmelzpunkt an mit sinkender Temperatur schnell ab und
  wird bei Metallen bei Metallen bei 1/3 der absoluten Schmelztemperatur
  unmerklich." which translates to ``The number of changes of place in the unit
  of time decreases rapidly from the melting point with falling temperature and
  becomes imperceptible for metals at 1/3 of the absolute melting
  temperature."\relax
\mciteBstWouldAddEndPuncttrue
\mciteSetBstMidEndSepPunct{\mcitedefaultmidpunct}
{\mcitedefaultendpunct}{\mcitedefaultseppunct}\relax
\EndOfBibitem
\bibitem[Becker and Dronskowski(2016)Becker, and Dronskowski]{becker2016first}
Becker,~N.; Dronskowski,~R. A first-principles study on new high-pressure
  metastable polymorphs of MoO2. \emph{Journal of Solid State Chemistry}
  \textbf{2016}, \emph{237}, 404--410, DOI:
  \doi{10.1016/j.jssc.2016.03.002}\relax
\mciteBstWouldAddEndPuncttrue
\mciteSetBstMidEndSepPunct{\mcitedefaultmidpunct}
{\mcitedefaultendpunct}{\mcitedefaultseppunct}\relax
\EndOfBibitem
\bibitem[Berger and Christophi(2003)Berger, and
  Christophi]{berger2003randomization}
Berger,~V.~W.; Christophi,~C.~A. Randomization technique, allocation
  concealment, masking, and susceptibility of trials to selection bias.
  \emph{Journal of Modern Applied Statistical Methods} \textbf{2003}, \emph{2},
  8, DOI: \doi{10.22237/jmasm/1051747680}\relax
\mciteBstWouldAddEndPuncttrue
\mciteSetBstMidEndSepPunct{\mcitedefaultmidpunct}
{\mcitedefaultendpunct}{\mcitedefaultseppunct}\relax
\EndOfBibitem
\bibitem[Cosby \latin{et~al.}(2020)Cosby, Mattei, Wang, Li, Bechtold, Chapman,
  and Khalifah]{cosby2020salt}
Cosby,~M.~R.; Mattei,~G.~S.; Wang,~Y.; Li,~Z.; Bechtold,~N.; Chapman,~K.~W.;
  Khalifah,~P.~G. Salt effects on Li-ion exchange kinetics of Na2Mg2P3O9N:
  Systematic in situ synchrotron diffraction studies. \emph{The Journal of
  Physical Chemistry C} \textbf{2020}, \emph{124}, 6522--6527, DOI:
  \doi{10.1021/acs.jpcc.0c00067}\relax
\mciteBstWouldAddEndPuncttrue
\mciteSetBstMidEndSepPunct{\mcitedefaultmidpunct}
{\mcitedefaultendpunct}{\mcitedefaultseppunct}\relax
\EndOfBibitem
\bibitem[Szymanski \latin{et~al.}(2021)Szymanski, Zeng, Huo, Bartel, Kim, and
  Ceder]{szymanski2021toward}
Szymanski,~N.~J.; Zeng,~Y.; Huo,~H.; Bartel,~C.~J.; Kim,~H.; Ceder,~G. Toward
  autonomous design and synthesis of novel inorganic materials. \emph{Materials
  Horizons} \textbf{2021}, \emph{8}, 2169--2198, DOI:
  \doi{10.1039/D1MH00495F}\relax
\mciteBstWouldAddEndPuncttrue
\mciteSetBstMidEndSepPunct{\mcitedefaultmidpunct}
{\mcitedefaultendpunct}{\mcitedefaultseppunct}\relax
\EndOfBibitem
\bibitem[Kimmig \latin{et~al.}(2021)Kimmig, Zechel, and
  Schubert]{kimmig2021digital}
Kimmig,~J.; Zechel,~S.; Schubert,~U.~S. Digital Transformation in Materials
  Science: A Paradigm Change in Material's Development. \emph{Advanced
  Materials} \textbf{2021}, \emph{33}, 2004940, DOI:
  \doi{10.1002/adma.202004940}\relax
\mciteBstWouldAddEndPuncttrue
\mciteSetBstMidEndSepPunct{\mcitedefaultmidpunct}
{\mcitedefaultendpunct}{\mcitedefaultseppunct}\relax
\EndOfBibitem
\bibitem[Chen \latin{et~al.}(2018)Chen, Hou, Chen, Tang, Langner, Li, Stubhan,
  Levchuk, Gu, Osvet, \latin{et~al.} others]{chen2018exploring}
Chen,~S.; Hou,~Y.; Chen,~H.; Tang,~X.; Langner,~S.; Li,~N.; Stubhan,~T.;
  Levchuk,~I.; Gu,~E.; Osvet,~A., \latin{et~al.}  Exploring the stability of
  novel wide bandgap perovskites by a robot based high throughput approach.
  \emph{Advanced Energy Materials} \textbf{2018}, \emph{8}, 1701543, DOI:
  \doi{10.1002/aenm.201701543}\relax
\mciteBstWouldAddEndPuncttrue
\mciteSetBstMidEndSepPunct{\mcitedefaultmidpunct}
{\mcitedefaultendpunct}{\mcitedefaultseppunct}\relax
\EndOfBibitem
\bibitem[Ortiz \latin{et~al.}(2019)Ortiz, Adamczyk, Gordiz, Braden, and
  Toberer]{ortiz2019towards}
Ortiz,~B.~R.; Adamczyk,~J.~M.; Gordiz,~K.; Braden,~T.; Toberer,~E.~S. Towards
  the high-throughput synthesis of bulk materials: thermoelectric
  PbTe--PbSe--SnTe--SnSe alloys. \emph{Molecular Systems Design \& Engineering}
  \textbf{2019}, \emph{4}, 407--420, DOI: \doi{10.1039/C8ME00073E}\relax
\mciteBstWouldAddEndPuncttrue
\mciteSetBstMidEndSepPunct{\mcitedefaultmidpunct}
{\mcitedefaultendpunct}{\mcitedefaultseppunct}\relax
\EndOfBibitem
\bibitem[Weston \latin{et~al.}(2019)Weston, Tshitoyan, Dagdelen, Kononova,
  Trewartha, Persson, Ceder, and Jain]{weston2019named}
Weston,~L.; Tshitoyan,~V.; Dagdelen,~J.; Kononova,~O.; Trewartha,~A.;
  Persson,~K.~A.; Ceder,~G.; Jain,~A. Named entity recognition and
  normalization applied to large-scale information extraction from the
  materials science literature. \emph{Journal of chemical information and
  modeling} \textbf{2019}, \emph{59}, 3692--3702, DOI:
  \doi{10.1021/acs.jcim.9b00470}\relax
\mciteBstWouldAddEndPuncttrue
\mciteSetBstMidEndSepPunct{\mcitedefaultmidpunct}
{\mcitedefaultendpunct}{\mcitedefaultseppunct}\relax
\EndOfBibitem
\bibitem[Trewartha \latin{et~al.}(2022)Trewartha, Walker, Huo, Lee, Cruse,
  Dagdelen, Dunn, Persson, Ceder, and Jain]{walker2021impact}
Trewartha,~A.; Walker,~N.; Huo,~H.; Lee,~S.; Cruse,~K.; Dagdelen,~J.; Dunn,~A.;
  Persson,~K.~A.; Ceder,~G.; Jain,~A. Quantifying the advantage of
  domain-specific pre-training on named entity recognition tasks in materials
  science. \emph{Patterns} \textbf{2022}, \emph{3}, 100488, DOI:
  \doi{10.1016/j.patter.2022.100488}\relax
\mciteBstWouldAddEndPuncttrue
\mciteSetBstMidEndSepPunct{\mcitedefaultmidpunct}
{\mcitedefaultendpunct}{\mcitedefaultseppunct}\relax
\EndOfBibitem
\bibitem[Nelder and Wedderburn(1972)Nelder, and
  Wedderburn]{nelder1972generalized}
Nelder,~J.~A.; Wedderburn,~R.~W. Generalized linear models. \emph{Journal of
  the Royal Statistical Society: Series A (General)} \textbf{1972}, \emph{135},
  370--384, DOI: \doi{10.2307/2344614}\relax
\mciteBstWouldAddEndPuncttrue
\mciteSetBstMidEndSepPunct{\mcitedefaultmidpunct}
{\mcitedefaultendpunct}{\mcitedefaultseppunct}\relax
\EndOfBibitem
\bibitem[Jain \latin{et~al.}(2013)Jain, Ong, Hautier, Chen, Richards, Dacek,
  Cholia, Gunter, Skinner, Ceder, \latin{et~al.} others]{jain2013commentary}
Jain,~A.; Ong,~S.~P.; Hautier,~G.; Chen,~W.; Richards,~W.~D.; Dacek,~S.;
  Cholia,~S.; Gunter,~D.; Skinner,~D.; Ceder,~G., \latin{et~al.}  Commentary:
  The Materials Project: A materials genome approach to accelerating materials
  innovation. \emph{APL materials} \textbf{2013}, \emph{1}, 011002, DOI:
  \doi{10.1063/1.4812323}\relax
\mciteBstWouldAddEndPuncttrue
\mciteSetBstMidEndSepPunct{\mcitedefaultmidpunct}
{\mcitedefaultendpunct}{\mcitedefaultseppunct}\relax
\EndOfBibitem
\bibitem[Wang \latin{et~al.}(2022)Wang, Kononova, Cruse, He, Huo, Fei, Zeng,
  Sun, Cai, Sun, \latin{et~al.} others]{wang2021dataset}
Wang,~Z.; Kononova,~O.; Cruse,~K.; He,~T.; Huo,~H.; Fei,~Y.; Zeng,~Y.; Sun,~Y.;
  Cai,~Z.; Sun,~W., \latin{et~al.}  Dataset of solution-based inorganic
  materials synthesis procedures extracted from the scientific literature.
  \emph{Scientific Data} \textbf{2022}, \emph{9}, 231, DOI:
  \doi{10.1038/s41597-022-01317-2}\relax
\mciteBstWouldAddEndPuncttrue
\mciteSetBstMidEndSepPunct{\mcitedefaultmidpunct}
{\mcitedefaultendpunct}{\mcitedefaultseppunct}\relax
\EndOfBibitem
\bibitem[Cruse \latin{et~al.}(2022)Cruse, Trewartha, Lee, Wang, Huo, He,
  Kononova, Jain, and Ceder]{Cruse2021}
Cruse,~K.; Trewartha,~A.; Lee,~S.; Wang,~Z.; Huo,~H.; He,~T.; Kononova,~O.;
  Jain,~A.; Ceder,~G. Text-mined dataset of gold nanoparticle synthesis
  procedures, morphologies, and size entities. \emph{Scientific Data}
  \textbf{2022}, \emph{9}, 234, DOI: \doi{10.1038/s41597-022-01321-6}\relax
\mciteBstWouldAddEndPuncttrue
\mciteSetBstMidEndSepPunct{\mcitedefaultmidpunct}
{\mcitedefaultendpunct}{\mcitedefaultseppunct}\relax
\EndOfBibitem
\bibitem[Bartel \latin{et~al.}(2018)Bartel, Millican, Deml, Rumptz, Tumas,
  Weimer, Lany, Stevanovi{\'c}, Musgrave, and Holder]{bartel2018physical}
Bartel,~C.~J.; Millican,~S.~L.; Deml,~A.~M.; Rumptz,~J.~R.; Tumas,~W.;
  Weimer,~A.~W.; Lany,~S.; Stevanovi{\'c},~V.; Musgrave,~C.~B.; Holder,~A.~M.
  Physical descriptor for the Gibbs energy of inorganic crystalline solids and
  temperature-dependent materials chemistry. \emph{Nature communications}
  \textbf{2018}, \emph{9}, 4168, DOI: \doi{10.1038/s41467-018-06682-4}\relax
\mciteBstWouldAddEndPuncttrue
\mciteSetBstMidEndSepPunct{\mcitedefaultmidpunct}
{\mcitedefaultendpunct}{\mcitedefaultseppunct}\relax
\EndOfBibitem
\bibitem[Bartel \latin{et~al.}(2020)Bartel, Trewartha, Wang, Dunn, Jain, and
  Ceder]{bartel2020critical}
Bartel,~C.~J.; Trewartha,~A.; Wang,~Q.; Dunn,~A.; Jain,~A.; Ceder,~G. A
  critical examination of compound stability predictions from machine-learned
  formation energies. \emph{npj Computational Materials} \textbf{2020},
  \emph{6}, 97, DOI: \doi{10.1038/s41524-020-00362-y}\relax
\mciteBstWouldAddEndPuncttrue
\mciteSetBstMidEndSepPunct{\mcitedefaultmidpunct}
{\mcitedefaultendpunct}{\mcitedefaultseppunct}\relax
\EndOfBibitem
\bibitem[Bartel \latin{et~al.}(2019)Bartel, Weimer, Lany, Musgrave, and
  Holder]{bartel2019role}
Bartel,~C.~J.; Weimer,~A.~W.; Lany,~S.; Musgrave,~C.~B.; Holder,~A.~M. The role
  of decomposition reactions in assessing first-principles predictions of solid
  stability. \emph{npj Computational Materials} \textbf{2019}, \emph{5}, 4,
  DOI: \doi{10.1038/s41524-018-0143-2}\relax
\mciteBstWouldAddEndPuncttrue
\mciteSetBstMidEndSepPunct{\mcitedefaultmidpunct}
{\mcitedefaultendpunct}{\mcitedefaultseppunct}\relax
\EndOfBibitem
\bibitem[Bartel(2022)]{bartel2022review}
Bartel,~C.~J. Review of computational approaches to predict the thermodynamic
  stability of inorganic solids. \emph{Journal of Materials Science}
  \textbf{2022}, \emph{57}, 10475--10498, DOI:
  \doi{10.1007/s10853-022-06915-4}\relax
\mciteBstWouldAddEndPuncttrue
\mciteSetBstMidEndSepPunct{\mcitedefaultmidpunct}
{\mcitedefaultendpunct}{\mcitedefaultseppunct}\relax
\EndOfBibitem
\bibitem[Not()]{Note-2}
NIST Chemistry WebBook.
  \href{https://webbook.nist.gov/chemistry/}{https://webbook.nist.gov/chemistry/}
  (accessed 2022-07-16).\relax
\mciteBstWouldAddEndPunctfalse
\mciteSetBstMidEndSepPunct{\mcitedefaultmidpunct}
{}{\mcitedefaultseppunct}\relax
\EndOfBibitem
\bibitem[Kim \latin{et~al.}(2021)Kim, Chen, Cheng, Gindulyte, He, He, Li,
  Shoemaker, Thiessen, Yu, \latin{et~al.} others]{kim2021pubchem}
Kim,~S.; Chen,~J.; Cheng,~T.; Gindulyte,~A.; He,~J.; He,~S.; Li,~Q.;
  Shoemaker,~B.~A.; Thiessen,~P.~A.; Yu,~B., \latin{et~al.}  PubChem in 2021:
  new data content and improved web interfaces. \emph{Nucleic acids research}
  \textbf{2021}, \emph{49}, D1388--D1395, DOI: \doi{10.1093/nar/gkaa971}\relax
\mciteBstWouldAddEndPuncttrue
\mciteSetBstMidEndSepPunct{\mcitedefaultmidpunct}
{\mcitedefaultendpunct}{\mcitedefaultseppunct}\relax
\EndOfBibitem
\bibitem[Not()]{Note-3}
FREED–Thermodynamic Database.
  \href{https://www.thermart.net/freed-thermodynamic-database/}{https://www.thermart.net/freed-thermodynamic-database/}
  (accessed 2022-07-16)\relax
\mciteBstWouldAddEndPuncttrue
\mciteSetBstMidEndSepPunct{\mcitedefaultmidpunct}
{\mcitedefaultendpunct}{\mcitedefaultseppunct}\relax
\EndOfBibitem
\bibitem[Pedregosa \latin{et~al.}(2011)Pedregosa, Varoquaux, Gramfort, Michel,
  Thirion, Grisel, Blondel, Prettenhofer, Weiss, Dubourg, Vanderplas, Passos,
  Cournapeau, Brucher, Perrot, and Duchesnay]{scikit-learn}
Pedregosa,~F. \latin{et~al.}  Scikit-learn: Machine Learning in {P}ython.
  \emph{Journal of Machine Learning Research} \textbf{2011}, \emph{12},
  2825--2830\relax
\mciteBstWouldAddEndPuncttrue
\mciteSetBstMidEndSepPunct{\mcitedefaultmidpunct}
{\mcitedefaultendpunct}{\mcitedefaultseppunct}\relax
\EndOfBibitem
\bibitem[Efron(1978)]{efron1978regression}
Efron,~B. Regression and ANOVA with zero-one data: Measures of residual
  variation. \emph{Journal of the American Statistical Association}
  \textbf{1978}, \emph{73}, 113--121, DOI:
  \doi{10.1080/01621459.1978.10480013}\relax
\mciteBstWouldAddEndPuncttrue
\mciteSetBstMidEndSepPunct{\mcitedefaultmidpunct}
{\mcitedefaultendpunct}{\mcitedefaultseppunct}\relax
\EndOfBibitem
\end{mcitethebibliography}

\end{document}